\documentclass[12pt]{article}

\usepackage[margin=1in]{geometry}  
\usepackage{graphicx}              
\usepackage{float}
\usepackage{caption}
\usepackage{subcaption}
\usepackage{amsmath}               
\usepackage{amsfonts}              
\usepackage{amsthm}                
\usepackage{algorithm}
\usepackage[noend]{algpseudocode}

\begin{document}

\title{\textbf{M}ultiple \textbf{A}lignment of \textbf{S}tructures using \textbf{C}enter \textbf{O}f Pro\textbf{T}eins}

\author{Kaushik Roy and Satish Ch. Panigrahi and Asish Mukhopadhyay\thanks{Research supported by an NSERC Discovery Grant to this author.} \\ \\ 
School of Computer Science \\
University of Windsor \\
401 Sunset Avenue \\
Windsor, ON N9B 2R4, Canada\\
\texttt{roy113@uwindsor.ca} \\
\texttt{panigra@uwindsor.ca} \\
\texttt{asishm@cs.uwindsor.ca}}

\maketitle

\section{Introduction}

Protein structure comparison has long been under investigation by computational 
biologists in hopes for finding a better but not necessarily faster alternative to the 
more familiar multiple sequence alignment problem. The multiple alignment problem is 
more challenging than pairwise alignment even for sequences, and we resort to 
heuristics to find as best an approximation as possible, in polynomial time. Since 
officially, as of 2014, there are more protein structures in the Protein Data Bank\cite{
Berman01012000} than seconds in a day, there is a constant need for both speed and 
precision when aligning more than two proteins. 

Multiple Structure Alignment (MStA) of protein structures can be categorized into four 
widely different approaches - progressive alignment, core optimization, graph based, 
and pivot based. 
Mustang \cite{konagurthu2006mustang}, 
msTali \cite{shealy2012multiple}
, mulPBA \cite{leonard2014mulpba}, and CE-MC \cite{guda2004mc} use the progressive 
alignment approach that creates an alignment of alignments following a guide tree. 
While this approach does make sense, it suffers from the natural disadvantages of all 
progressive techniques. Methods from other approaches, 
\cite{menke2008matt}
\cite{ye2005multiple}, often outperform the progressive ones, both in terms of speed and accuracy. 
A second way is to optimize a consensus structure, sometimes with several iterations, 
and report a common core of the input proteins. The idea is to find out a structurally 
conserved subset of residues among the proteins to gain some insight into their origin. 
However, such cores are mostly pseudo-structures that, although geometrically 
interesting, may or may not have any biological relevance. 
Matt \cite{menke2008matt}, 
Multiprot \cite{shatsky2002multiprot}, 
Mass \cite{dror2003mass}, 
Mapsci \cite{ye2006multiple}, and 
Smolign \cite{sun2012smolign} belong to this category of MStA 
algorithms. Ye and Godzik's graph based POSA \cite{ye2005multiple} takes a totally 
different path by representing a protein as a directed acyclic graph (DAG) of residues 
connected in the order following the backbone. POSA then creates a combined non-planar 
multi-dimensional DAG, taking hinge rotation into account, to come up with residue 
equivalences among the input proteins. While POSA gains the upper hand in terms of 
flexibility of protein structures, it is known to completely miss motifs on TIM barrel 
and helix bundles proteins \cite{sun2012smolign} and incur higher cost of alignment 
than MATT and Smolign \cite{sun2012smolign}. The pivot based approach selects one of 
the input molecules as 'closest' to all the other proteins and names it the pivot. The 
remaining proteins are then iteratively aligned to the pivot either in a bottom-up 
manner \cite{wang2009fast}, or in a top-down manner \cite{ye2004approximate} to come up 
with residue-residue correspondences that are later used to minimize some objective 
function and derive a score as a similarity measure. Some of the only few published 
algorithms in this category, are Mistral \cite{micheletti2009mistral}, 
\cite{wang2009fast}, and \cite{ye2004approximate}. Our approach is an application of the center-star 
method of producing a multiple sequence alignment (MSA) in an MStA scenario.

In this paper we introduce a new algorithm, \textbf{MASCOT}, which stands for \textbf{M}ultiple \textbf{A}lignment of \textbf{S}tructures using \textbf{C}enter \textbf{O}f pro\textbf{T}eins, for aligning more than two proteins. Our algorithm relies on taking advantage of the linear nature of the protein polypeptide chain, while judiciously preserving the secondary structure elements (SSEs). The justification for this approach is that SSEs are latent ingredients of protein structures, serving as a well-preserved scaffold. As a result, SSEs are evolutionarily remarkably conserved while changes happen in the loops, thus modifying functionality, e.g. the substrate specificity of different serine proteases is governed by the conformation of the binding loops \cite{hedstrom2002serine}. Further, representing protein structures by their SSEs has been successfully used on several occasions in pairwise alignment (\cite{koch1996algorithm}, \cite{alesker1996detection}, \cite{alexandrov1996analysis}; \cite{grindley1993identification}, \cite{lu2000top}). Our goal was to develop an algorithm that uses these sequences of SSEs and produces a multiple alignment with minimal running time and comparable accuracy. To this end, we have implemented MASCOT as a hybrid algorithm that uses a center protein, derived by minimizing sum of pairwise distance, to drive a layout that identifies a set of residues from each protein that are structurally similar. We then proceed to find an optimal correspondence among the backbone carbon atoms of these molecular structures, using inter-residue Euclidean distance threshold, and report the centerRMSD of the structures aligned in space as a measure of similarity.

\section{Method}
\label{sect:proposed}

\subsection{Input data set}
Protein structures are stored as PDB files in Protein Data Bank \cite{Berman01012000}, which contains more than 99000 structures and is regularly updated. The data can come from standard protein databases available, or from a local repository. Either way this section retrieves the correct data and makes it readily available for the algorithm to proceed. This may seem trivial at first, but for all practical purposes where thousands of macromolecules could be aligned together, one must make take care that all the requisite molecules are fetched and ready for the next step. As an example, an input set could be 1AOR:A 7ACN 2ACT 1TTQ:B etc.
As we can see, with inputs having such varied insignia, to supply the right data to the algorithm we need this first step. 
\subsection{Representing the proteins}
Once we are assured access to the correct input set, we take each protein in turn and represent them is a way which makes further processing convenient while keeping all vital information intact.

This is a key step in the process of obtaining a useful multiple alignment due to the following factors:-

\begin{description}
  \item[(a)] Too simple a representation could potentially miss crucial structural and functional information, whereas too complex a representation will demand innovative methods at every turn; primary sequence is the simplest possible representation and its known that they do not necessarily determine functionality. POSA on the other hand uses partial order graphs, while mulPBA uses PROFIT elements.
  \item[(b)] Difference between equal and unequal length proteins; for equal length proteins we can represent the proteins by their coordinates and apply Umeyama's method\cite{panigrahi2014eigendecomposition}. Janardan's method\cite{ye2004approximate} handles unequal length proteins by representing the protein as a set of vectors using gap vectors for a gap alignment.
  \item[(c)] Choice of representation greatly affects performance;
  
\end{description}

Thus, to strike a balance between complexity and functionality, we represent the molecules as their DSSP sequences \cite{kabsch1983dssp}, which assigns each residue to one of eight possible structural motifs in a protein. Simply put, we use the linear polymer nature of the protein and stretch it out into a straight line keeping SSEs intact; which is sufficient for biological purposes. \\

For example a dssp representation of SEA CUCUMBER CAUDINA (1HLM) would look like this:
\begin{center}
...HHHHGGGZZIIIITTHHHHHHTTSSI...
\end{center}

For N input proteins we get N such dssp representations.

The main advantage of this is that we are now free to use a host of pattern matching algorithms that can compute an optimal alignment given any two such sequences - an observation which we shall exploit in the next step.
\subsection{Pairwise global alignment}
The processing begins by applying global alignment between every pair of dssp sequences. To do this we apply Needleman Wunsch algorithm \cite{needleman1970general} with appropriate affine gap opening penalty, and the following scoring matrix: \\

\begin{tabular}{cc}
    \begin{minipage}{.5\linewidth}
        \begin{tabular}{|c|c|c|c|c|c|c|c|c|}
         \hline
         $-$ & H & B & G & E & T & S & I & Z \\
         \hline
         H & 1 & 0 & 1 & 0 & 0 & 0 & 1 & 0\\
         \hline
         B & 0 & 1 & 0 & 1 & 0 & 0 & 0 & 0\\
         \hline
         G & 1 & 0 & 1 & 0 & 0 & 0 & 1 & 0\\
         \hline
         E & 0 & 1 & 0 & 1 & 0 & 0 & 0 & 0\\
         \hline
         T & 0 & 0 & 0 & 0 & 1 & 1 & 0 & 0\\
         \hline
         S & 0 & 0 & 0 & 0 & 1 & 1 & 0 & 0\\
         \hline
         I & 1 & 0 & 1 & 0 & 0 & 0 & 1 & 0\\
         \hline
         Z & 0 & 0 & 0 & 0 & 0 & 0 & 0 & 1\\
         \hline
         \end{tabular}
    \end{minipage} &

\begin{minipage}{.5\linewidth}
\begin{tabular}{|c||l|}
\hline
$Symbol$ & $Motif$ \\
\hline
H & Alpha Helix \\
\hline
B & Beta bridge \\
\hline
G & Helix 3 \\
\hline
E & Beta strand \\
\hline
T & Turn \\
\hline
S & Bend \\
\hline
I & Helix 5 \\
\hline
Z & No motif \\
\hline
\end{tabular}
\end{minipage} 
\end{tabular}
\\ \\ \\
These pairwise alignments produce a primitive picture of SSE SSE alignments. For example an alignment between dssp sequences of globins 1DM1, 1MBC, 1MBA produce the following output. 

\begin{table}[h]
\begin{center}
\begin{tabular}{c l}

 \hspace{1mm}  \hspace{1mm} & \hspace{1mm}  \hspace{1mm}  \\ 
  
\texttt{1DM1} & \texttt{..ZZZHHHHHHHHHHHHHHHHTHHHHHHHHHHHHHHHSGGG-...}   \\ 
\texttt{1MBA} & \texttt{..ZZZHHHHHHHHHHHHHHHHT-HHHHHHHHHHHHHHHZGGG...}  \\ 
 &    \\ 
\texttt{1DM1} & \texttt{..ZZZHHHHHHHHHHHHHHHHTHHHHHHHHHHHHHHH-SGGG...}   \\ 
\texttt{1MBC} & \texttt{..ZZZHHHHHHHHHHHHHHGGGHHHHHHHHHHHHHHHZTHHH...}  \\ 
 &   \\ 
\texttt{1MBC} & \texttt{..ZZZHHHHHHHHHHHHHHGGGHHHHHHHHHHHHHHHZTHH-H...}   \\ 
\texttt{1MBA} & \texttt{..ZZZHHHHHHHHHHHHHHHHTHHHHHHHHHHHHHHHZGGGGG...}  \\ 
\end{tabular}
\end{center}
\end{table}

We can see that the helices are properly aligned against one another. These alignments are saved in a list and referred to when needed. 
\subsection{Center protein}
Keeping with the center star method \cite{gusfield1993efficient} for multiple sequence alignment, an NxN symmetric matrix is created where the entries are edit distances between every aligned pair in the list mentioned above. 
\begin{center}
\begin{tabular}{|c|c|c|c|c|}
\hline
$-$ & $P_1$ & $P_2$ & \dots & $P_N$\\
\hline
$P_1$ & 0 & 10 & \dots & 20\\
\hline
$P_2$ & 10 & 0 & \dots & 30\\
\hline
\vdots & $\cdots$ & $\cdots$ & $\cdots$ & \vdots\\
\hline
$P_N$ & 20 & 30 & $\cdots$ & 0\\
\hline
\end{tabular}
\end{center}

From this matrix we find the sum of pair score and the center protein using the following equation. 
\begin{equation}
P_{c} \text{ = protein with minimum }\sum_{i\not=j}^N EditDistance(P_{i},P_{j})
\end{equation}

The protein having the minimum sum of pair score is chosen as the center protein with respect to which other proteins can be aligned. A pairwise edit distance matrix for globins 1DM1, 1MBC, 1MBA is given below:

\begin{table}[h]
\caption{\label{Comparison table}Pairwise Distance matrix} 
\begin{center}

\begin{tabular}{c c c c}
   
 \hspace{1mm}  \hspace{1mm} & \hspace{1mm} 1DM1 \hspace{1mm} & \hspace{1mm} 1MBC \hspace{1mm} & \hspace{1mm} 1MBA \hspace{1mm}\\ 
  
1DM1 & 0 &  37 &  4  \\ 
1MBC & 34 &  0 &  35  \\ 
1MBA & 7 & 35 &  0   \\ 
\end{tabular}

\end{center}

\end{table}

The protein, at index c, having the minimum SP distance is labelled as the center protein P$_c$ and its dssp sequence as S$_c$. Among the three proteins above 1DM1 is selected as the center protein since it has the lowest SP distance (equal to 41).

A lot of heuristics go on to calculate a consensus structure in place of a real protein, in hopes of finding a common core. Mustang, MultiProt, Janaradan, all attempt to procure a template structure to drive their alignment process. However, it should be noted that a consensus structure, while geometrically and perhaps computationally convenient, can often turn out to be a pseudo structure that may or may not be of any actual biological significance. Thus, our choice of going along with an actual protein to accelerate the alignment, can be deemed appropriate.
\subsection{Correspondence matrix}
Formally put, a correspondence matrix is an N x $l$ matrix with respect to a center protein $P_{c}$ where, \\ \\
N=$|P| , P = \{P_{1}, P_{2},\dots, P_{N}\}$ being the set of proteins, and \\ \\
max$(|P_{1}|, |P_{2}|,\dots,|P_{N}|) \le l \le |P_{1}|+|P_{2}|+\dots+|P_{N}|$ \\ \\
A correspondence matrix using the above guidelines will have the following properties:
\begin{description}
  \item[(a)] The $i^{th}$ row contains the ordered set of residues from protein Pi, with gaps in between.
  \item[(b)] No column will have all gaps
  \item[(c)] Gives a good idea of residue equivalences to work with, should we have to apply rigid body superposition.
  
\end{description}
In this step, all the alignments, between the center protein and every other protein, are retrieved from the list and merged one by one using  the following algorithm: \\ 
\begin{algorithm}
    \caption*{\textbf{Algorithm:} \textsc{Correspondence Matrix}} \label{alg:dfs1}

    \begin{algorithmic}
        \Require Protein DSSP sequences $S_{1}$, $S_{2}$ upto $S_{N}$
        \Ensure MSA of Sequences $S_{1}$ to $S_{N}$
        \Statex
        \ForAll{$i=1...N-1$}
        	\State use alignment $(S_c,S_i)$ and MSA$(S_c, S_{i-1})$ to obtain MSA$(S_c, S_{i})$ followng 'once a gap, always a gap' rule
        \EndFor

    \end{algorithmic}
\end{algorithm}\\

A sample correspondence matrix for the globin family is given below. We can clearly see how the SSEs of all the proteins are aligned together in a column-wise fashion. \\

\begin{figure}[h]
\begin{center}
\includegraphics[width=\textwidth]{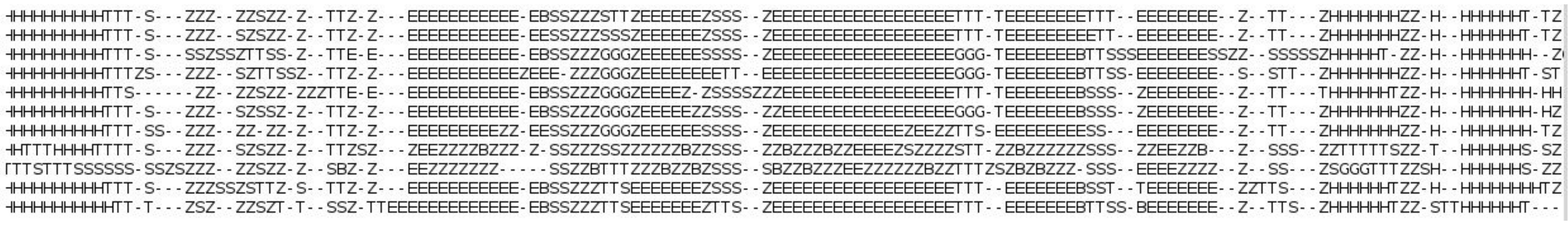}
\end{center}
\caption{A correspondence matrix. [Notice there are no columns with gaps in all rows.]}
\label{fig:doodle}
\end{figure}

At this point we have identified conserved regions across all the proteins, but not aligned them in any way. Janaradan's method reaches a similar result, creating a correspondence matrix by carefully manipulating vectors.

 The result is an MSA of dssp sequences S$_i$ for proteins P$_i$, 1 $\leq$ i $\leq$ N. The output of this step is used to identify as many residue equivalences as possible given the raw protein structures. To actually align them in 3D space we feed this output to the next step.

\subsection{Rigid body superposition}
To align the proteins in space means to apply proper translation and rotation so that the distance between alpha carbon atoms of equivalent residues is below a threshold value. To do this we need two things; a set of equivalences, and a reference frame against which the rigid body superposition is to take place. We already have both these requirements fulfilled. The correspondence matrix from phase 2 gives us the residue-residue equivalences, and our chosen center protein is the reference frame. So, suppose the correspondence is as below:

\begin{table}[H]
\caption{\label{equivalences}Identifying equivalences} 
\begin{center}
\begin{tabular}{l||c c c c c c c c c}
   
Residue no. &  &  & 1 & 2 & 3 & 4 & 5 &  & 6\\ 
  
Center protein & -  & - & H & H & T & I & E & - & G\\ 
Other protein & S & S & H & H & - & G & E & E & I\\ 
Residue no. & 1 & 2 & 3 & 4 &  & 5 & 6 & 7 & 8\\ 
\end{tabular}
\end{center}

\end{table}
Some annotated equivalences between center and the other protein would be (1,3) (2,4) (4,5) (5,6) and (6,8).

For each protein $P_i$ we apply Kabsch's method \cite{kabsch1976solution} to superpose the structures, in space,with respect to the center protein $P_c$. For example, our algorithm aligns globins 1DM1, 1MBC, 1MBA in this way. 

\begin{figure}[!htb]
\minipage{0.32\textwidth}
	\includegraphics[width=\linewidth]{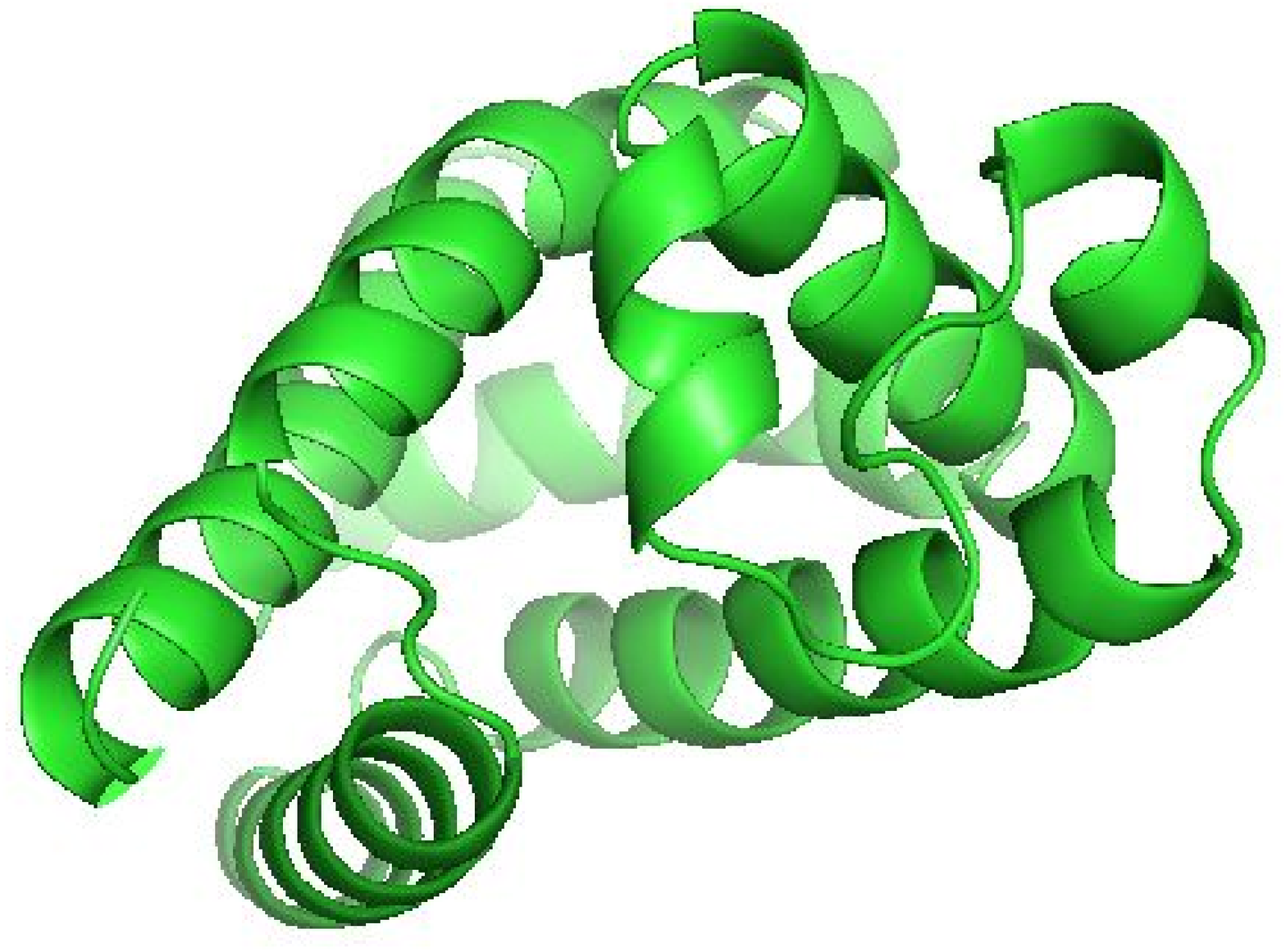}
  \caption{1DM1}\label{}
\endminipage\hfill
\minipage{0.32\textwidth}
	\includegraphics[width=\linewidth]{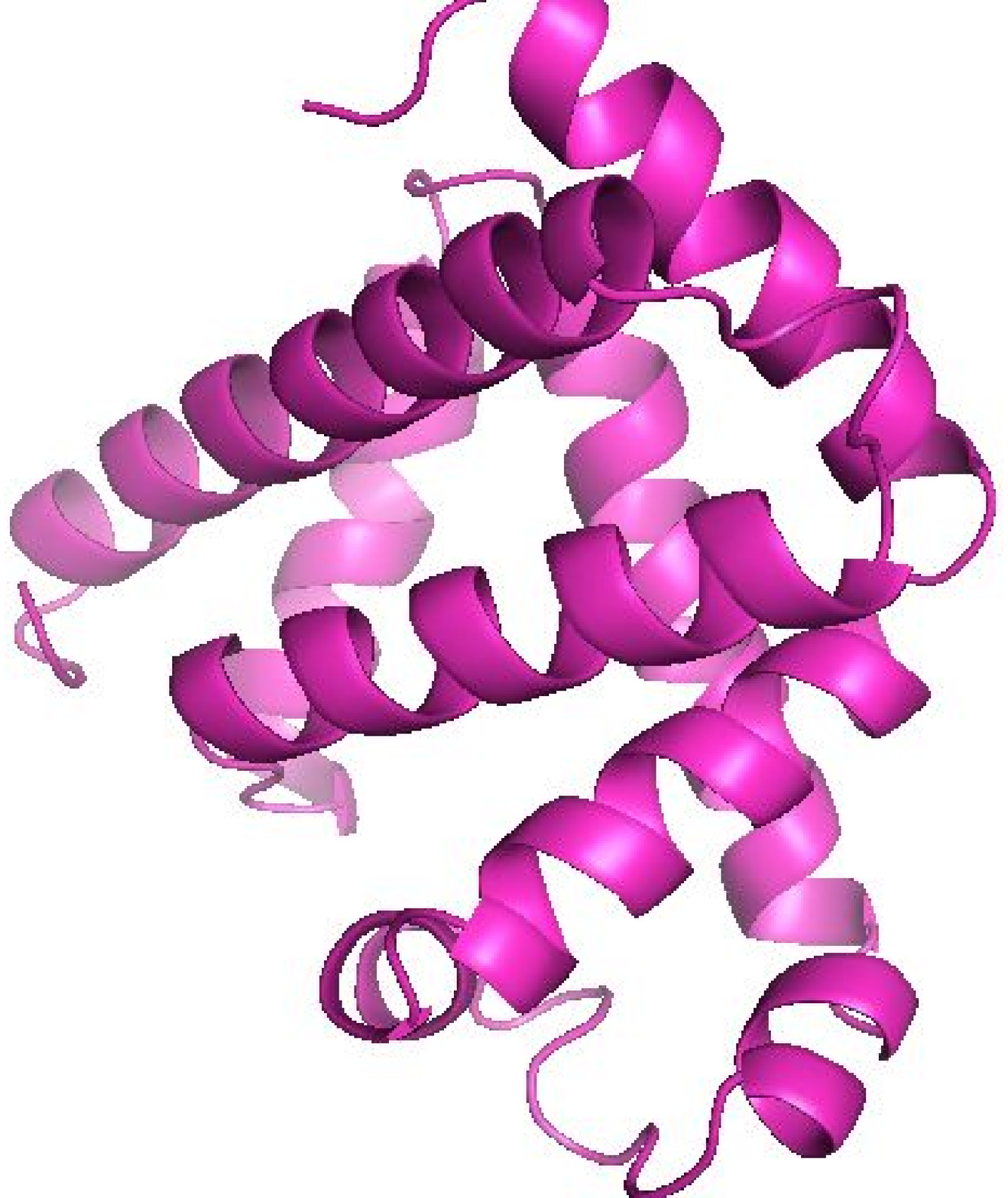}
  \caption{1MBC}\label{}
\endminipage\hfill
\minipage{0.32\textwidth}%
	\includegraphics[width=\linewidth]{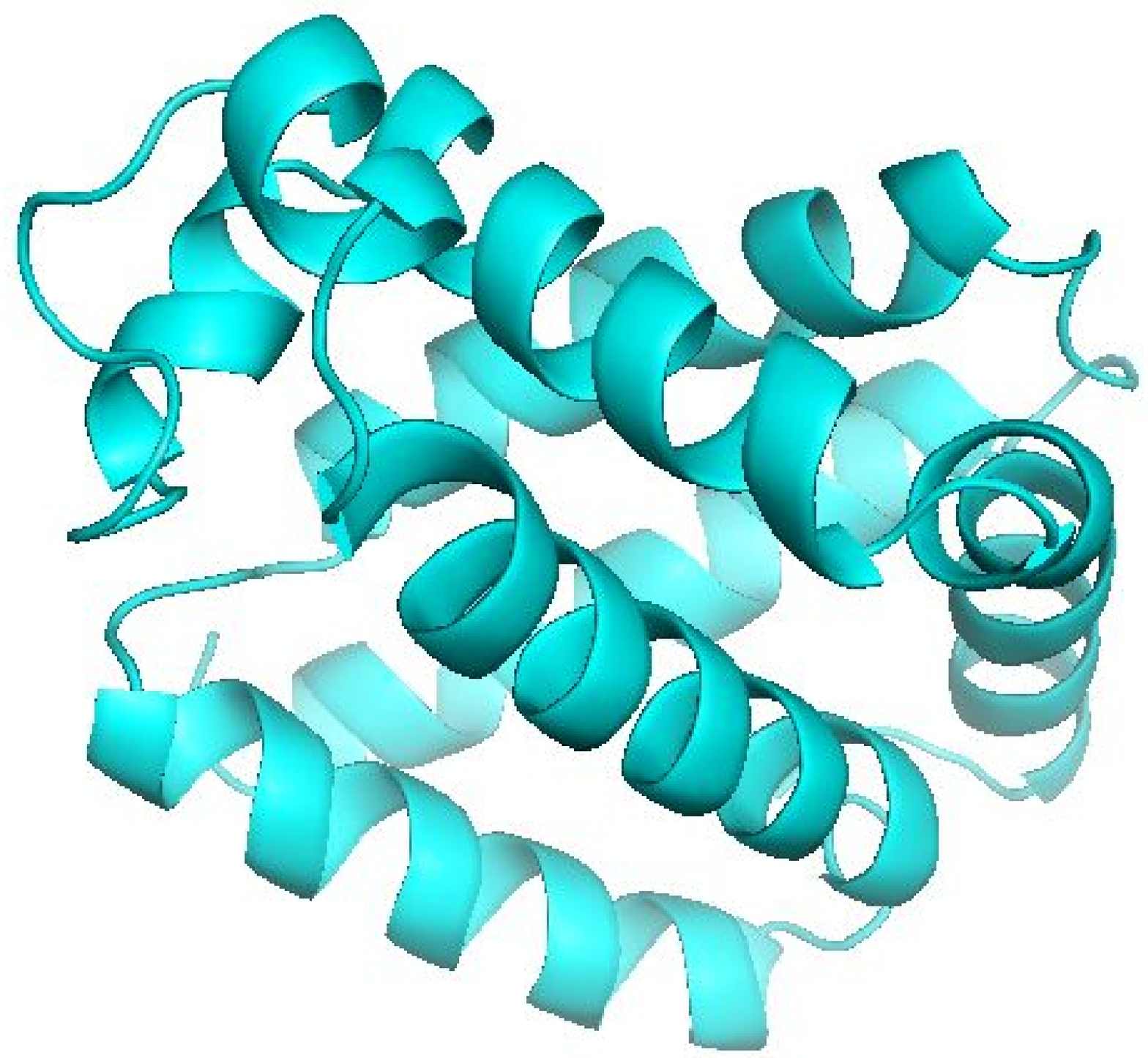}
  \caption{1MBA}\label{}
\endminipage
\end{figure}
\begin{figure}[h]
\makebox[\linewidth]{
\includegraphics[scale=0.7]{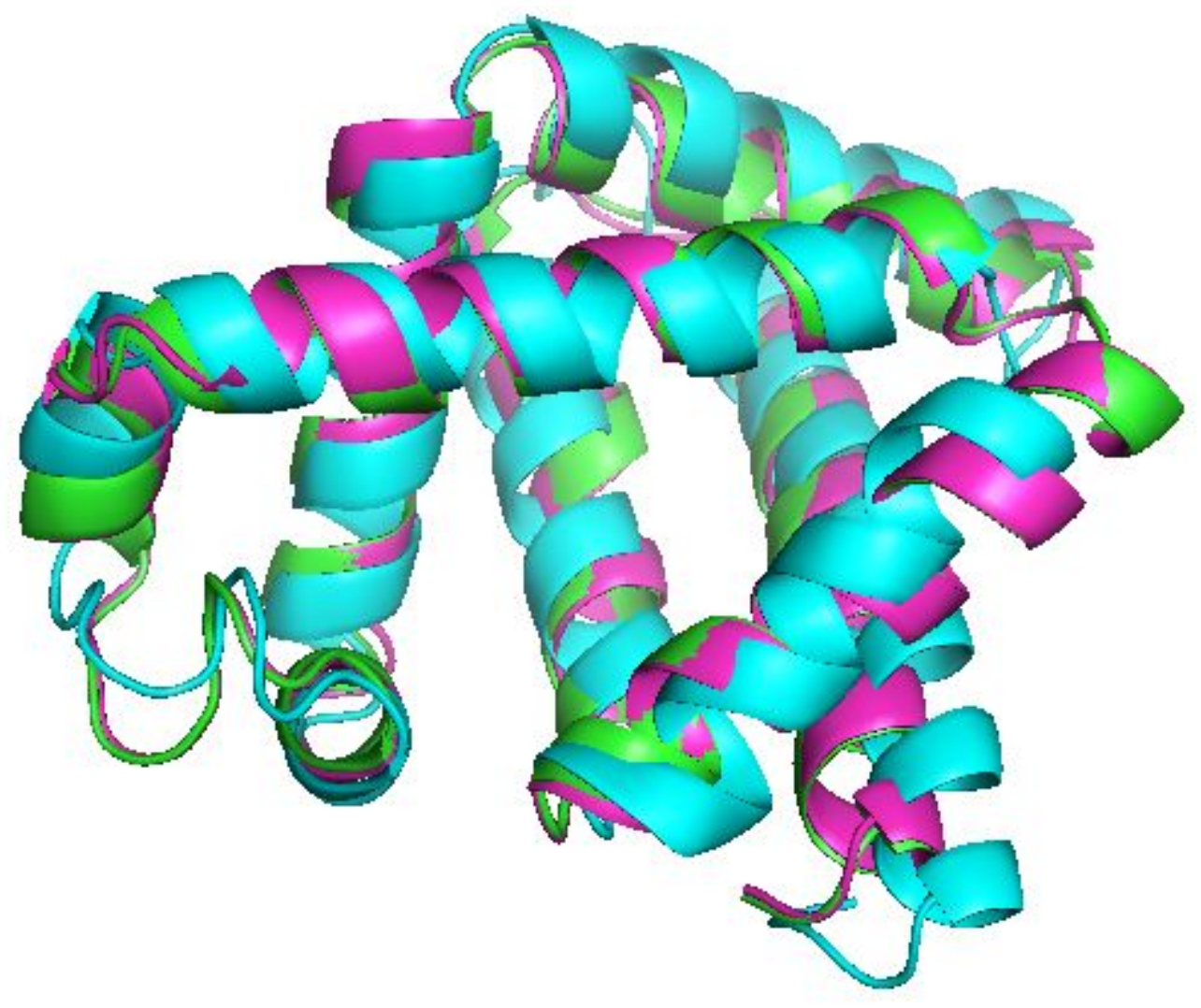}
}
\caption{Alignment of 1dm1, 1mbc, and 1mba}
\end{figure}
The picture above clearly shows how the structures are aligned in three dimensions.
\subsection{Dynamic programming and scoring}
Once the proteins are aligned in space through rigid body superposition, equivalent residues have been brought close to each other. We increase then number of equivalences between a pair of proteins by calculating the Euclidean distance between every pair of alpha carbon atoms, and considering the ones that fall within a threshold value(5\AA{}), as equivalent pairs.

Finally we use the following formulae to derive the \textit{centerRMSD} that represents the quality of the alignment. 
\[   \frac{1}{N-1} \sum_{i=1,i \neq c}^{N} RMSD(P_i,P_c)\]
A good alignment is one where the score is typically less than half the threshold value(2.5\AA{}). However, difficult alignments having biological relevance can exceed this value by about 1.5\AA{}. The centerRMSD for the alignment in Fig. 5 is \texttt{0.44}\AA{}.

\subsection{Pseudocode}
A pseudo-code form of our algorithm is as follows:

\begin{algorithm}
    \caption*{\textbf{Algorithm:} \textsc{MASCOT}} \label{alg:dfs1}

    \begin{algorithmic}[1]
        \Require Protein $pdbids (pdbid_1,pdbid_1,\cdots,pdbid_N)$
        \Ensure Multiple alignment of proteins with files created for $pdbids_{1...N}$
        \Statex 
        \Comment Phase 1
        \Statex
        \State Extract protein structures into P=\{$P_{1}$ to $P_{N}$\}
        \State Represent $P$ as sequences S=\{$S_{1}$ to $S_{N}$\} consisting of DSSP SSE elements 
        \State Perform pairwise global alignment of every $(S_i,S_j)$ using custom similarity matrix
        \Statex 
        \Comment Phase 2
        \Statex
        \State Create an edit distance matrix that stores the distances between every $(P_i,P_j)$ using a custom scoring function
        \State Choose the protein(sequence) with index $c$ having minimum \emph{sum of pairs} score as the center protein(sequence) $P_c$($S_c$)
        \State Create an MSA of $S$ w.r.t $S_c$ using center-star approach
        \Statex \Comment Phase 3
        \Statex
        \State Treat all alignments of symbols with non-gaps as residue-residue equivalences bet. $(P_i,P_j)$
        \State Apply Kabsch's method on every $(P_i,P_c)$ to obtain $(trans_i,rot_i)$ for every $(P_i,P_c)$ 
        \State Use $(trans_i,rot_i)$ from step 8 to transform and place $P_i$ in space, with $P_c$ being brought to origin first, to produce output pdb files 

    \end{algorithmic}
\end{algorithm}

\section{Results and Discussion}
\label{sect:results}
MASCOT was implemented in Python 2.7.5 using packages from Bio-python 2.0. A plethora of experiments were conducted, among which a representative set of results have been presented here. Note that T represents the time taken right from giving the input to producing the output files.

\subsection{Globins}
Globins are some of the most rigorously studied proteins by the MStA community. The globin family has long been known from studies of approximately 150-residue proteins such as vertebrate myoglobins and haemoglobins. The following globins have been aligned using MASCOT:

\begin{table}[h]
\caption{\label{Comparison table}The table below shows the globins used in this section:} 
\begin{center}

\begin{tabular}{|c|p{10cm}|c|c|}
\hline
   
 \hspace{1mm} Name \hspace{1mm} & \hspace{1mm} PDB ids \hspace{1mm} & \hspace{1mm} Count \hspace{1mm} & \hspace{1mm} T \hspace{1mm}\\ \hline
  
Set 1 & 1HHO:A 2DHB:A 2DHB:B 1HHO:B 1MBD 1DLW 1DLY 1ECO 1IDR:A 2LH7 &  10 &  23s  \\ \hline
Set 2 & 1MBC 1MBA 1DM1 1HLM 2LHB 2FAL 1HBG 1FLP 1ECA 1ASH &  10 &  24s  \\ \hline
Set 3 & 5MBN 1ECO 2HBG 2LH3 2LHB 4HHB:B 4HHB:A & 7 &  13s   \\ \hline
Set 4 & 1ASH 1ECA 1GDJ 1HLM 1MBA 1BAB:A 1EW6:A 1H97:A 1ITH:A 1SCT:A 1DLW:A 1FLP 1HBG 1LHS 1MBC 1DM1 2LHB 2FAL 1HBG 1FLP & 20 &  1m 38s   \\ \hline
\end{tabular}

\end{center}

\end{table}

\begin{figure}[!htb]
\minipage{0.25\textwidth}
	\includegraphics[width=\linewidth]{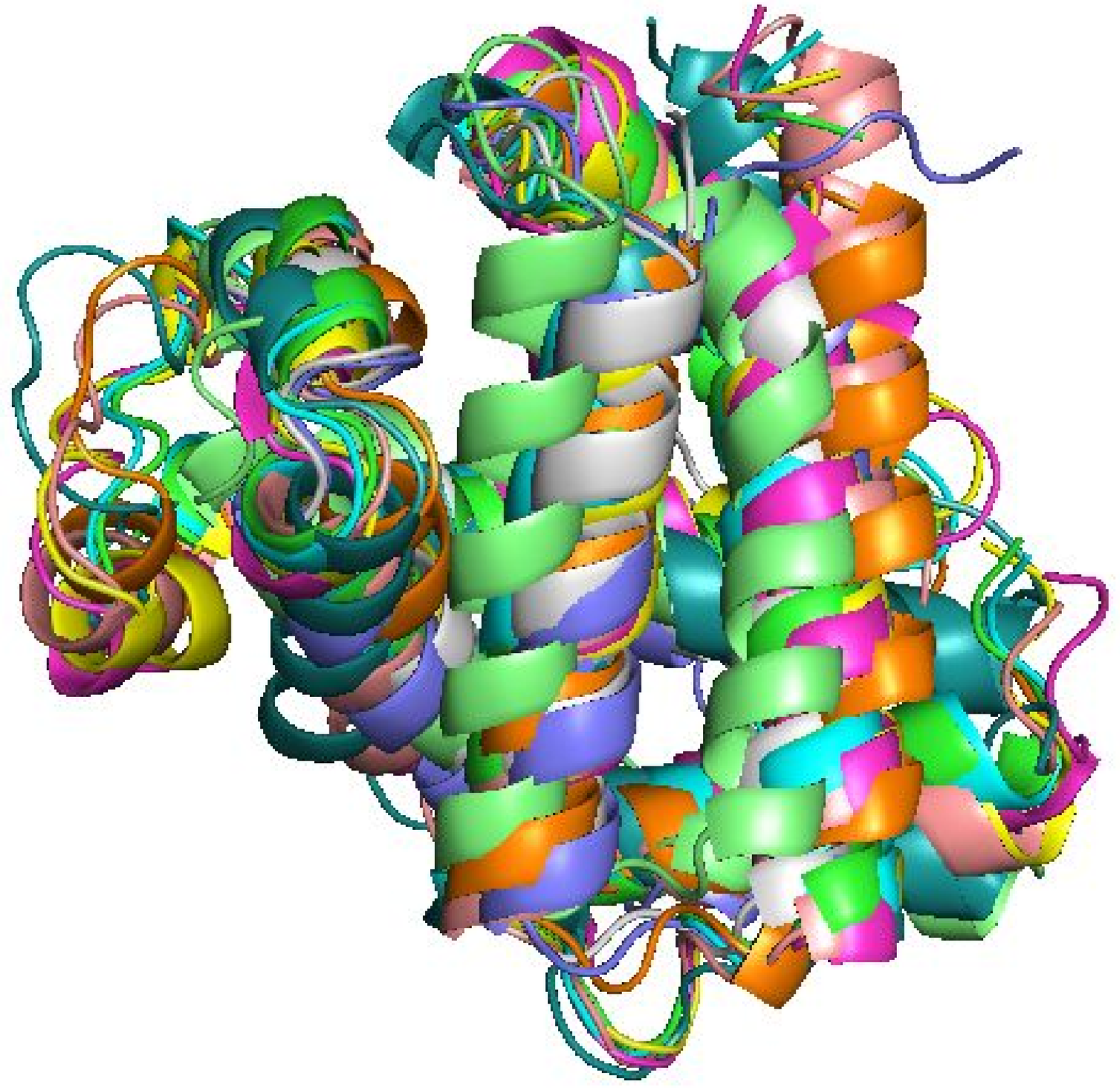}
  \caption{Set 1}\label{}
\endminipage\hfill
\minipage{0.25\textwidth}
	\includegraphics[width=\linewidth]{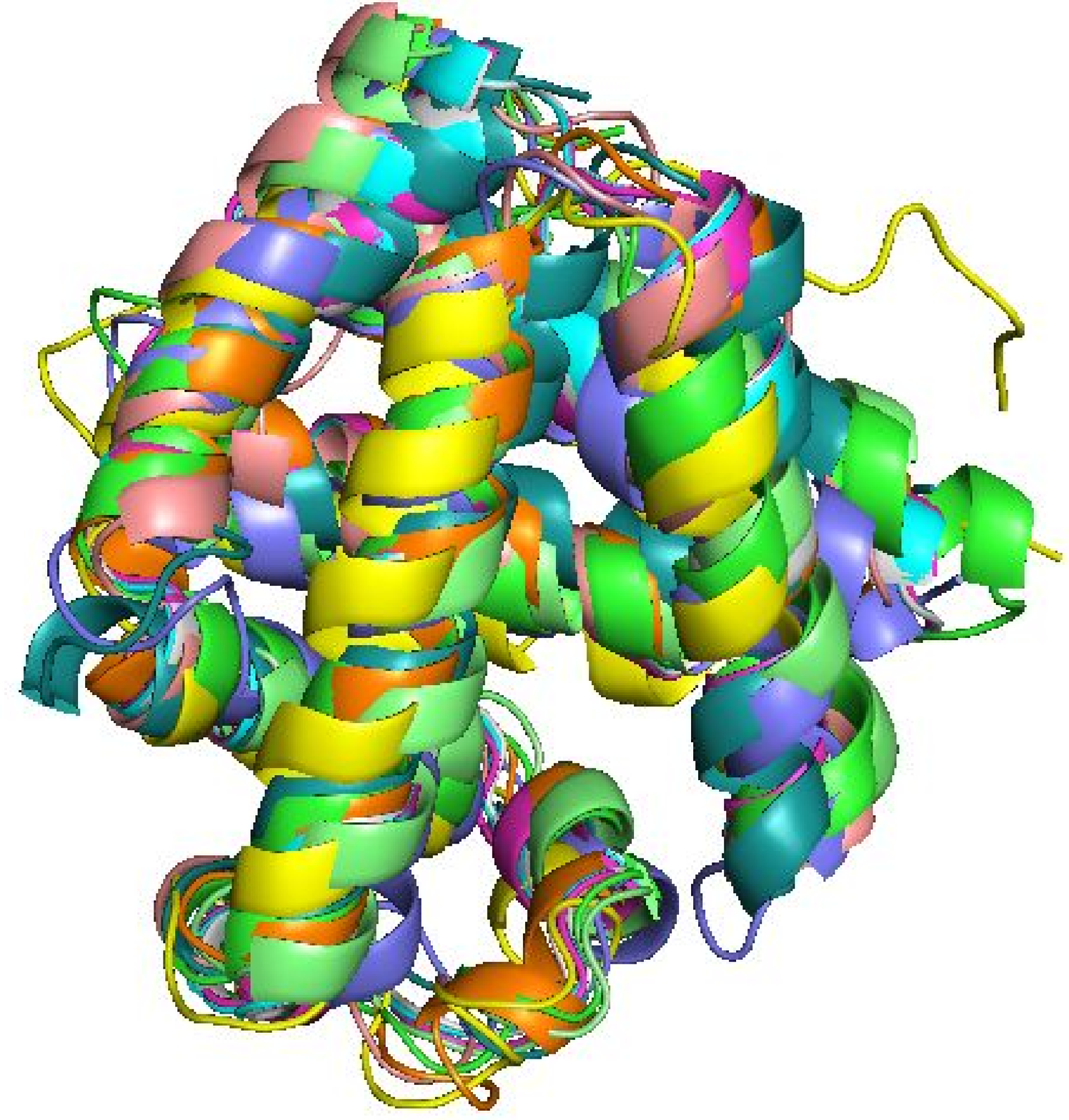}
  \caption{Set 2}\label{}
\endminipage\hfill
\minipage{0.25\textwidth}%
	\includegraphics[width=\linewidth]{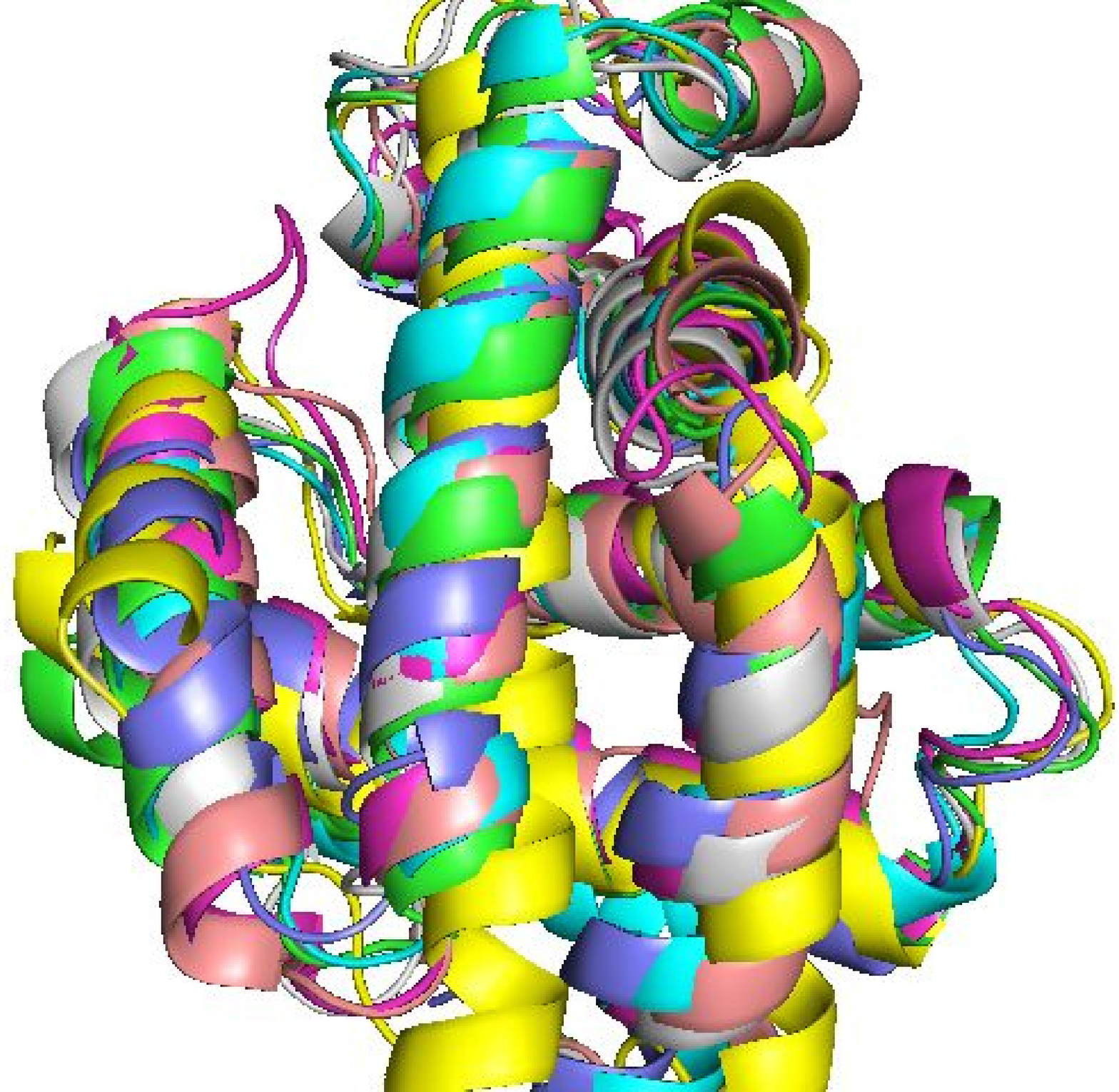}
  \caption{Set 3}\label{}
\endminipage
\minipage{0.25\textwidth}%
	\includegraphics[width=\linewidth]{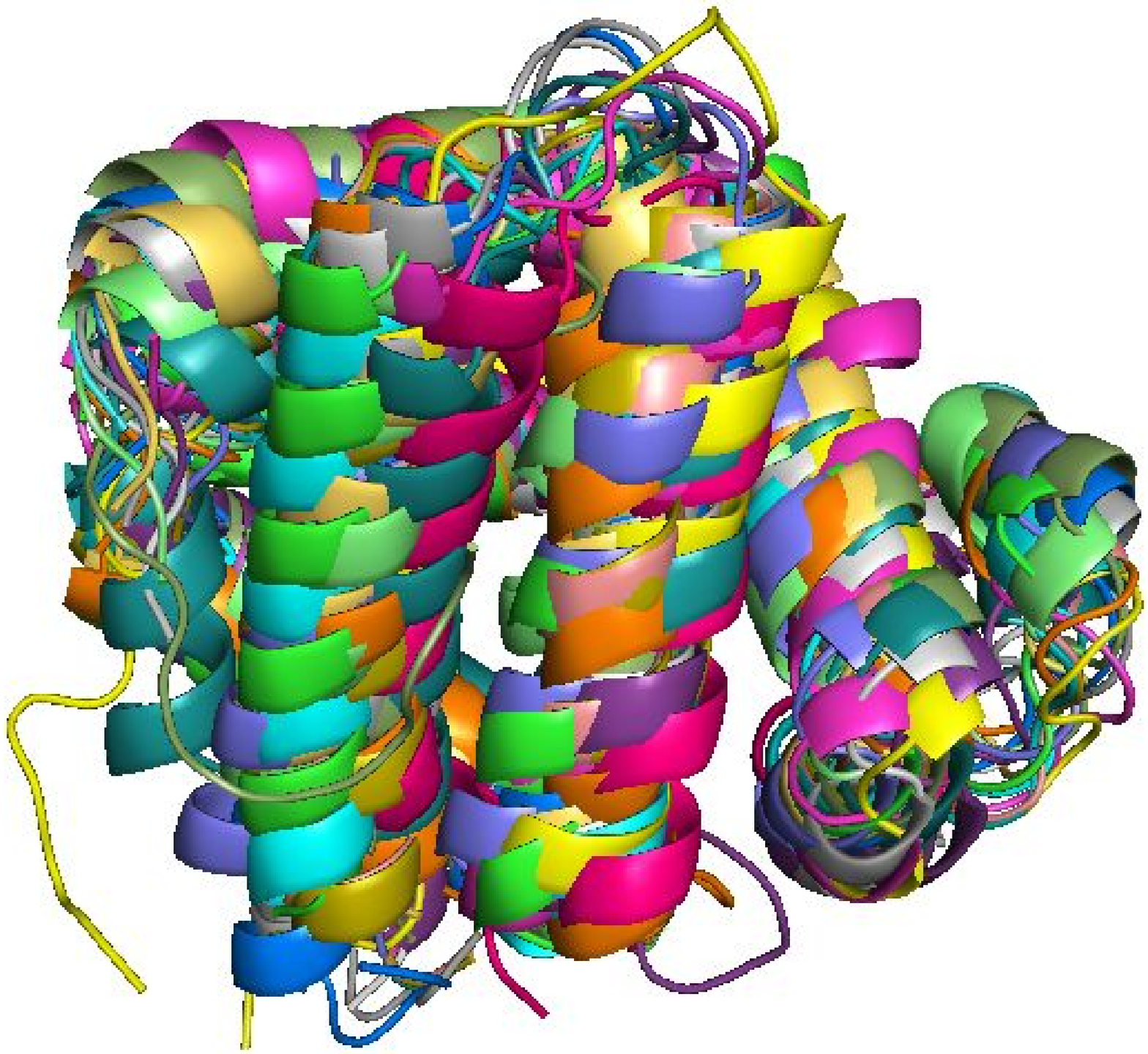}
  \caption{Set 4}\label{}
\endminipage
\end{figure}
Set 1 is used by \cite{micheletti2009mistral}, and \cite{sun2012smolign} to show how their algorithms align globins. The rmsd for this superposition is \texttt{2.765}\AA{}. Set 2, taken from \cite{ye2004approximate}, has been aligned with an rmsd of \texttt{2.39}\AA{}. Set 3 is \cite{shatsky2002multiprot}'s test data with rmsd \texttt{2.41}\AA{}. Set 4 is a custom assortment of 20 globins created from \cite{lupyan2005new} and \cite{ye2006multiple}. The purpose is to see how well they are aligned visually and with how much rmsd. As one can see, the helices and the hinges are placed within the threshold distance as much as possible, with rmsd \texttt{2.038}\AA{}.
\subsection{Serpins}
Serpins play an important role in the biological world. For instance, thyroxine-hinding globulin is a serpine which transports hormones to various parts of the body, and Maspin is a serpine which controls gene expression of certain tumors \cite{bernardo2011maspin}. The name Serpin stands for Serine Protease Inhibitors. The following serpins have been aligned using MASCOT:
\begin{table}[h]
\caption{\label{Comparison table}The table below shows the serpins used in this section:} 
\begin{center}

\begin{tabular}{|c|p{10cm}|c|c|}
\hline
   
 \hspace{1mm} Name \hspace{1mm} & \hspace{1mm} PDB ids \hspace{1mm} & \hspace{1mm} Count \hspace{1mm} & \hspace{1mm} T \hspace{1mm}\\ \hline
  
Set 1 & 7API:A 8API:A 1HLE:A 1OVA:A 2ACH:A 9API:A 1PSI 1ATU 1KCT 1ATH:A 1ATT:A 1ANT:L 2ANT:L &  13 &  3m 33s  \\ \hline
\end{tabular}
\end{center}

\end{table}

\begin{figure}[!htb]
\minipage{0.50\textwidth}
	\includegraphics[width=\linewidth]{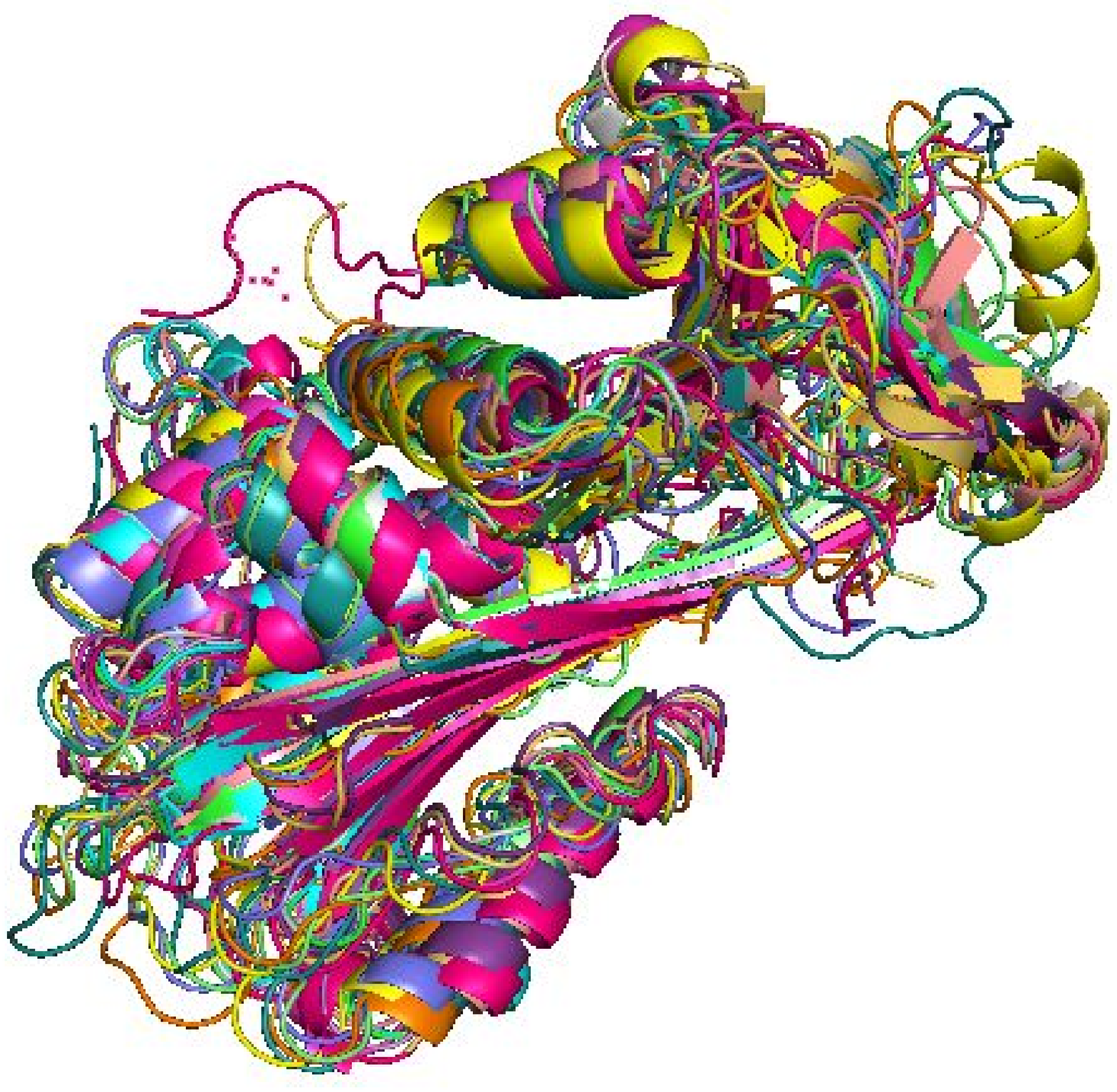}
  \caption{Set 5}\label{fig:awesome_image1}
\endminipage\hfill
\minipage{0.50\textwidth}
	\includegraphics[width=\linewidth]{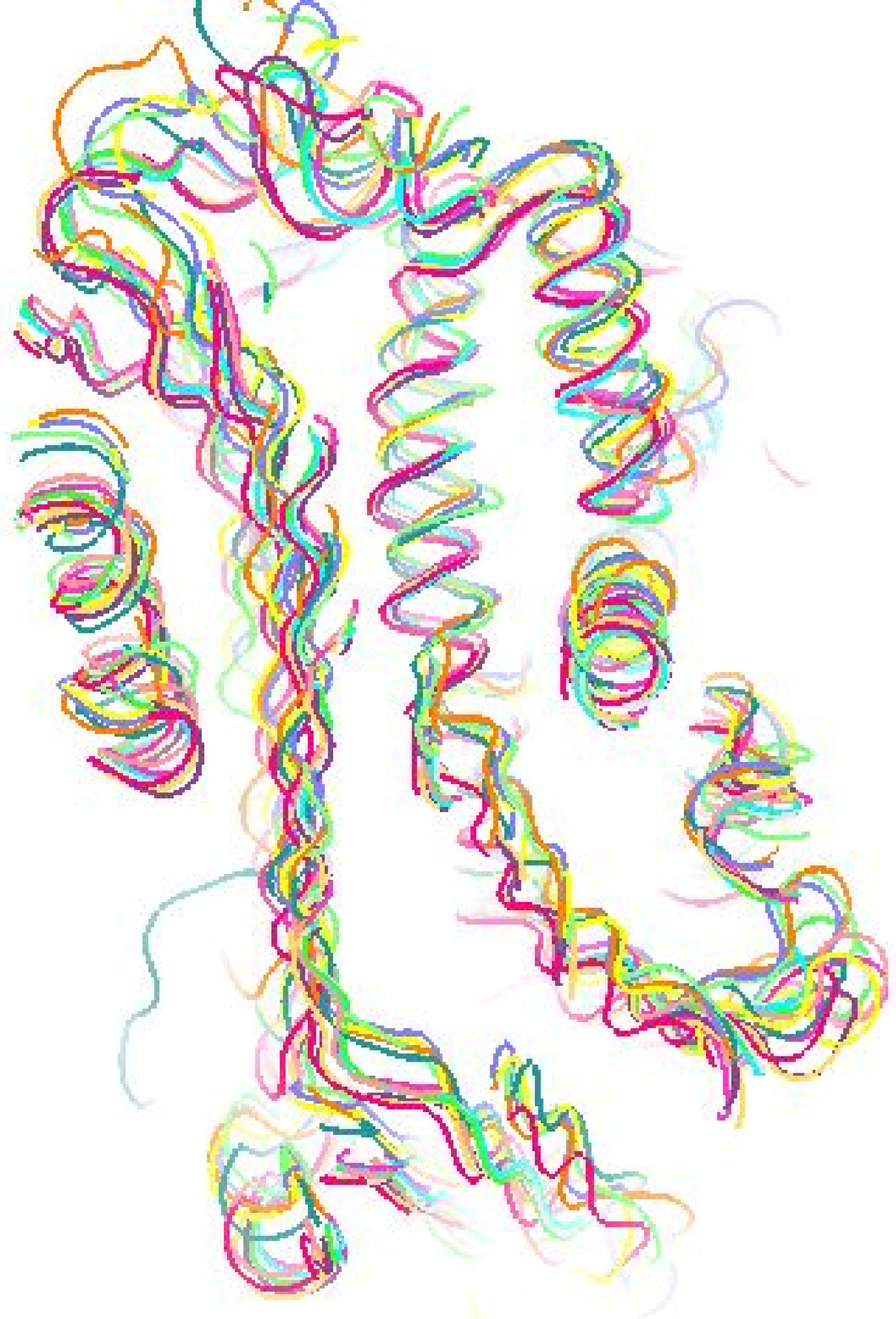}
  \caption{Set 5 LIPR}\label{fig:awesome_image2}
\endminipage\hfill
\end{figure}

The serpins in set 5 is the same one used by \cite{shatsky2002multiprot} and is said to be quite difficult owing to their large size and motif distribution. Unlike \cite{shatsky2002multiprot} we do not attempt to find a common core. Instead, we perform a global alignment over the length of the proteins. Fig. 10 shows how the beta sheets, hinges, and helices are aligned together in spite of the difficulty. Also some non-alignable parts have been correctly identified and left out. The rmsd for this alignment is \texttt{2.99}\AA{}. Fig. 11 is a low intensity PyMol rendition (LIPR) of the same alignment viewed from another angle. It uses a ribbon representation to condense the output and show most of the aligned portions of the proteins. The pictures suggest that all these serpins share functionality and purpose, within the body. We can club all these proteins into a single family, and keep adding to it as and when such high similarities are found.
\subsection{Barrels}
The eight-stranded TIM-barrel is found in a lot of enzymes, but the evolutionary history of this family has been the subject of rigorous debate. The ancestry of this family is still a mystery. Aligning TIM-barrel proteins will allow us to add to this ever-expanding family. The proteins aligned in this category are as follows:
\begin{table}[h]
\caption{\label{Comparison table}The table below shows the barrels used in this section:} 
\begin{center}
\begin{tabular}{|c|p{10cm}|c|c|}
\hline
   
 \hspace{1mm} Name \hspace{1mm} & \hspace{1mm} PDB ids \hspace{1mm} & \hspace{1mm} Count \hspace{1mm} & \hspace{1mm} T \hspace{1mm}\\ \hline
  
Set 6 & 1A49:A 1A49:B 1A49:C 1A49:D 1A49:E 1A49:F 1A49:G 1A49:H 1A5U:A 1A5U:B 1A5U:C 1A5U:D 1A5U:E 1A5U:F 1A5U:G 1A5U:H 1AQF:A 1AQF:B 1AQF:C 1AQF:D 1AQF:E 1AQF:F 1AQF:G 1AQF:H 1F3X:A 1F3X:B 1F3X:C 1F3X:D 1F3X:E 1F3X:F 1F3X:G 1F3X:H 1PKN 1F3W:A 1F3W:B 1F3W:C 1F3W:D 1F3W:E 1F3W:F 1F3W:G 1F3W:H 1PKM 1PKL:A 1PKL:B 1PKL:C 1PKL:D 1PKL:E 1PKL:F 1PKL:G6 1PKL:H 1A3W:A 1A3W:B 1A3X:A 1A3X:B 1E0T:A 1E0T:B 1E0T:C 1E0T:D 1PKY:A 1PKY:B 1PKY:C 1PKY:D7 1E0U:A 1E0U:B 1E0U:C 1E0U:D &  66 &  2h 25m  \\ \hline
Set 7 & 1SW3:A 1SW3:B 1WYI:A 1WYI:B 2JK2:A 2JK2:B 1R2T:A 1R2T:B 1R2R:A 1R2R:B 1M5W:A 1M5W:B 1M5W:C &  13 &  1m 22s  \\ \hline
\end{tabular}
\end{center}
\end{table}
\begin{figure}[!htb]
\minipage{0.32\textwidth}
	\includegraphics[width=\linewidth]{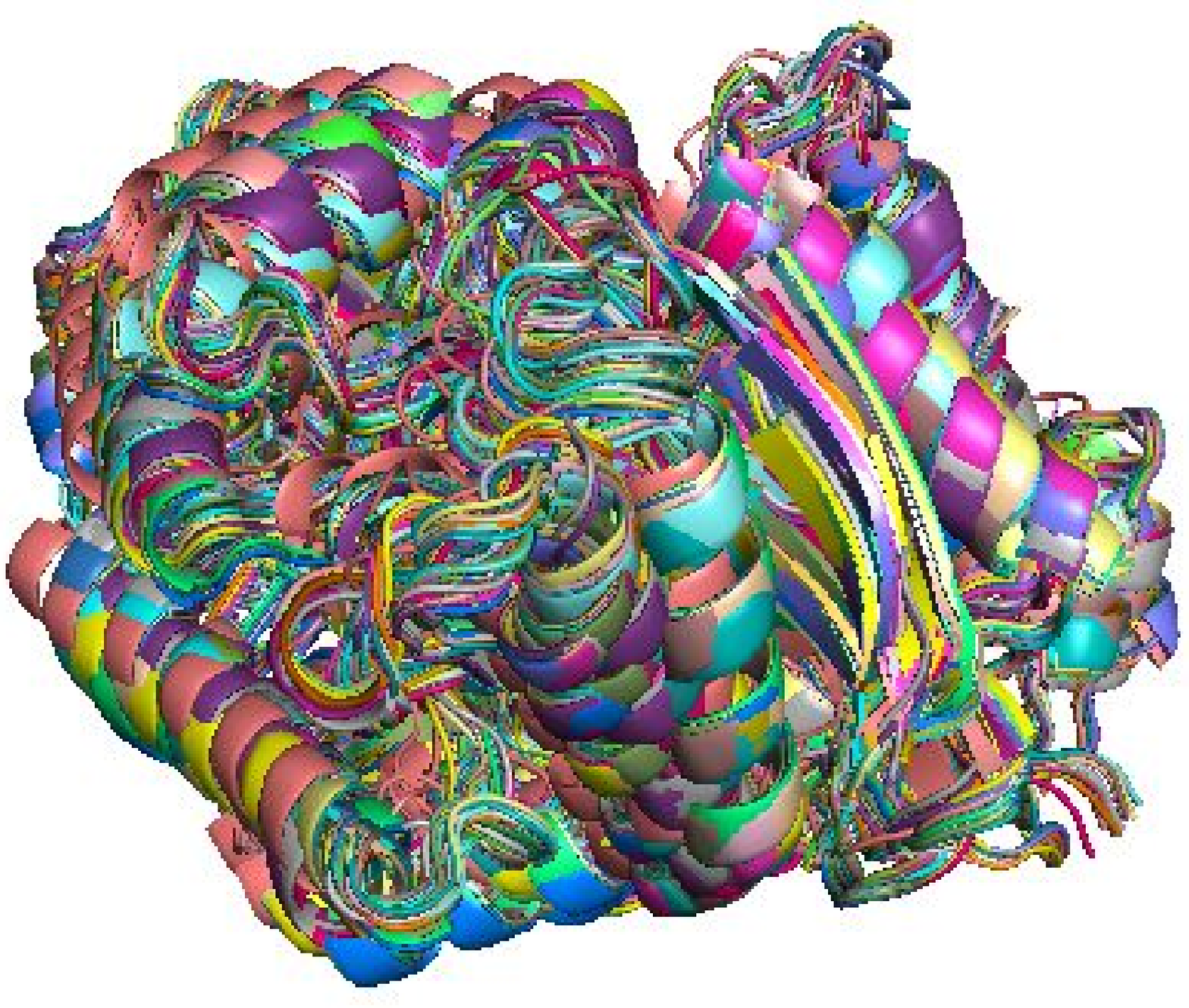}
  \caption{Set 6}\label{}
\endminipage\hfill
\minipage{0.32\textwidth}
	\includegraphics[width=\linewidth]{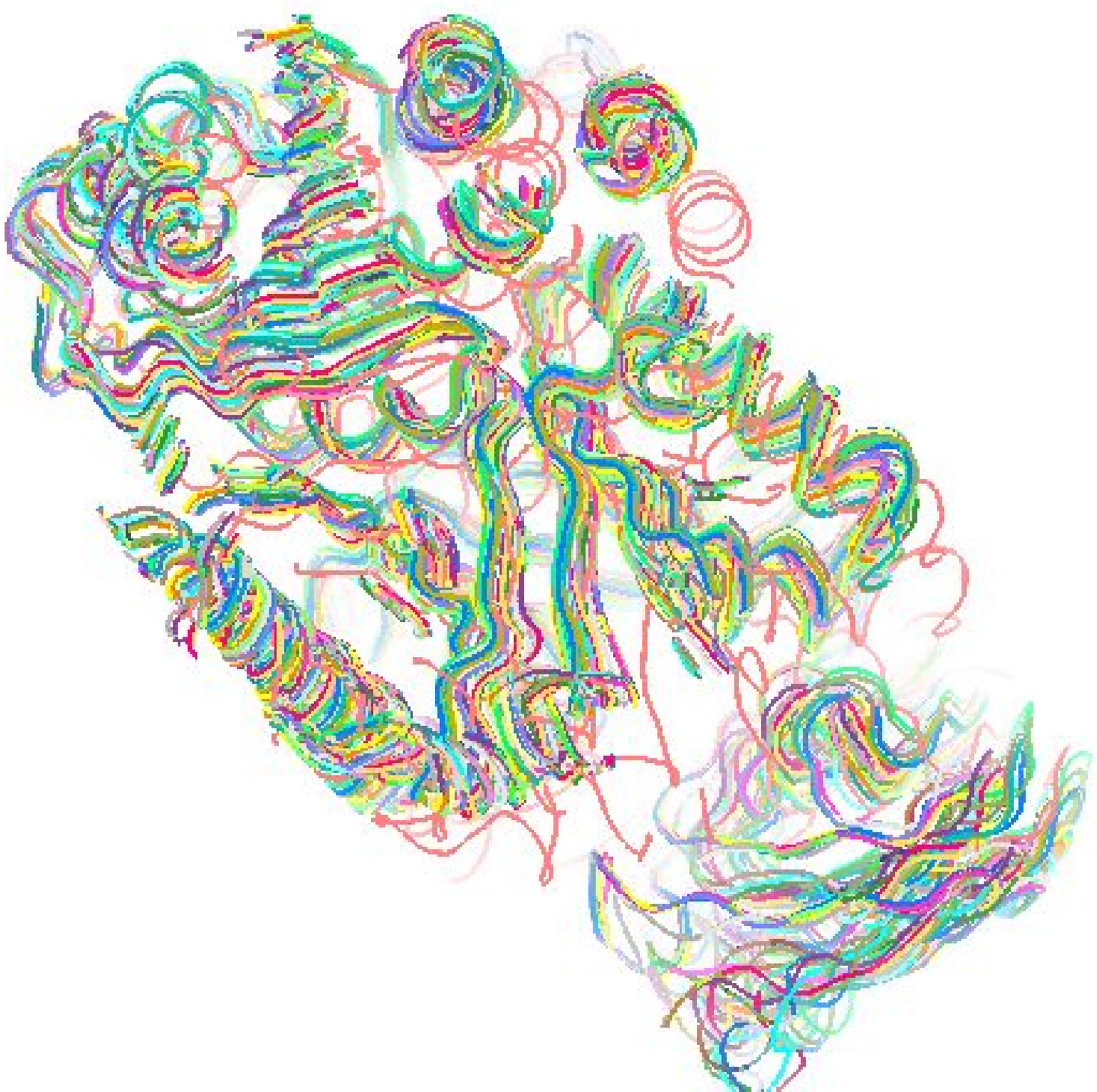}
  \caption{Set 6 LIPR}\label{}
\endminipage\hfill
\minipage{0.32\textwidth}
	\includegraphics[width=\linewidth]{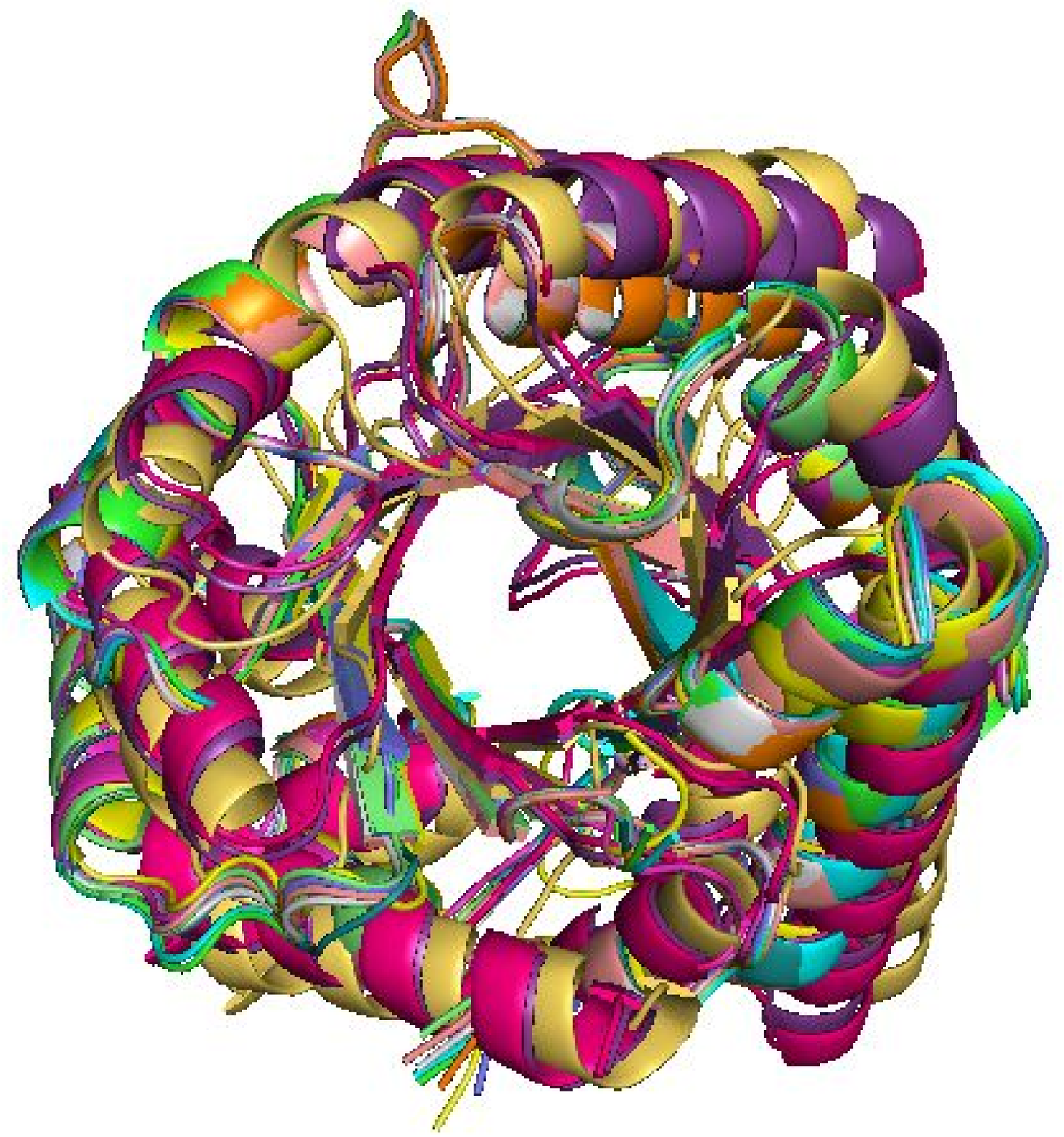}
  \caption{Set 7}\label{}
\endminipage\hfill
\end{figure}

MASS \cite{dror2003mass} has used the 66 molecules in set 6 to show how it aligns proteins with barrels. MASCOT produces an rmsd of \texttt{3.4}\AA{} for this alignment. Fig. 12 shows how the new algorithm can superimpose proteins having the TIM barrel supermotifs. Fig. 13 is an LIPR of the same alignment, for convenience. The result indicates these proteins have structurally highly conserved regions since all 8 helices and 8 beta sheets have been aligned. Set 7 has been taken from the gold standard manually curated SCOP database. The proteins are taken from different superfamilies but, as Fig. 14 suggests, MASCOT is still able to align the barrel motifs on top of each other, with an rmsd of \texttt{3.76}\AA{}.

\subsection{Twilight-zone proteins}
Sequence alignment is still an option except when proteins have less than 30\% sequence identity. The lesser the sequence similarity, the more important becomes structural comparison. Here we have taken some data sets that belong to the twilight zone.
\begin{table}[h]
\caption{\label{}The table below shows the sets used in this section:} 
\begin{center}
\begin{tabular}{|c|p{8cm}|c|c|}
\hline
   
 \hspace{1mm} Name \hspace{1mm} & \hspace{1mm} PDB ids \hspace{1mm} & \hspace{1mm} Seq. Identity \hspace{1mm} & \hspace{1mm} T \hspace{1mm}\\ \hline
  
Set 8 & 1STF:I 1MOL:A 1CEW:I &  \textless8\% &  1m 55s  \\ \hline
Set 9 & 1BGE:A 1BGE:B 2GMF:A 2GMF:B &  \textless12\% &  5s  \\ \hline
Set 10 & 1NSB 2SIM 1F8E 4DGR &  \textless20\% &  19s  \\ \hline
\end{tabular}
\end{center}
\end{table}
\begin{figure}[!htb]
\minipage{0.32\textwidth}
	\includegraphics[width=\linewidth]{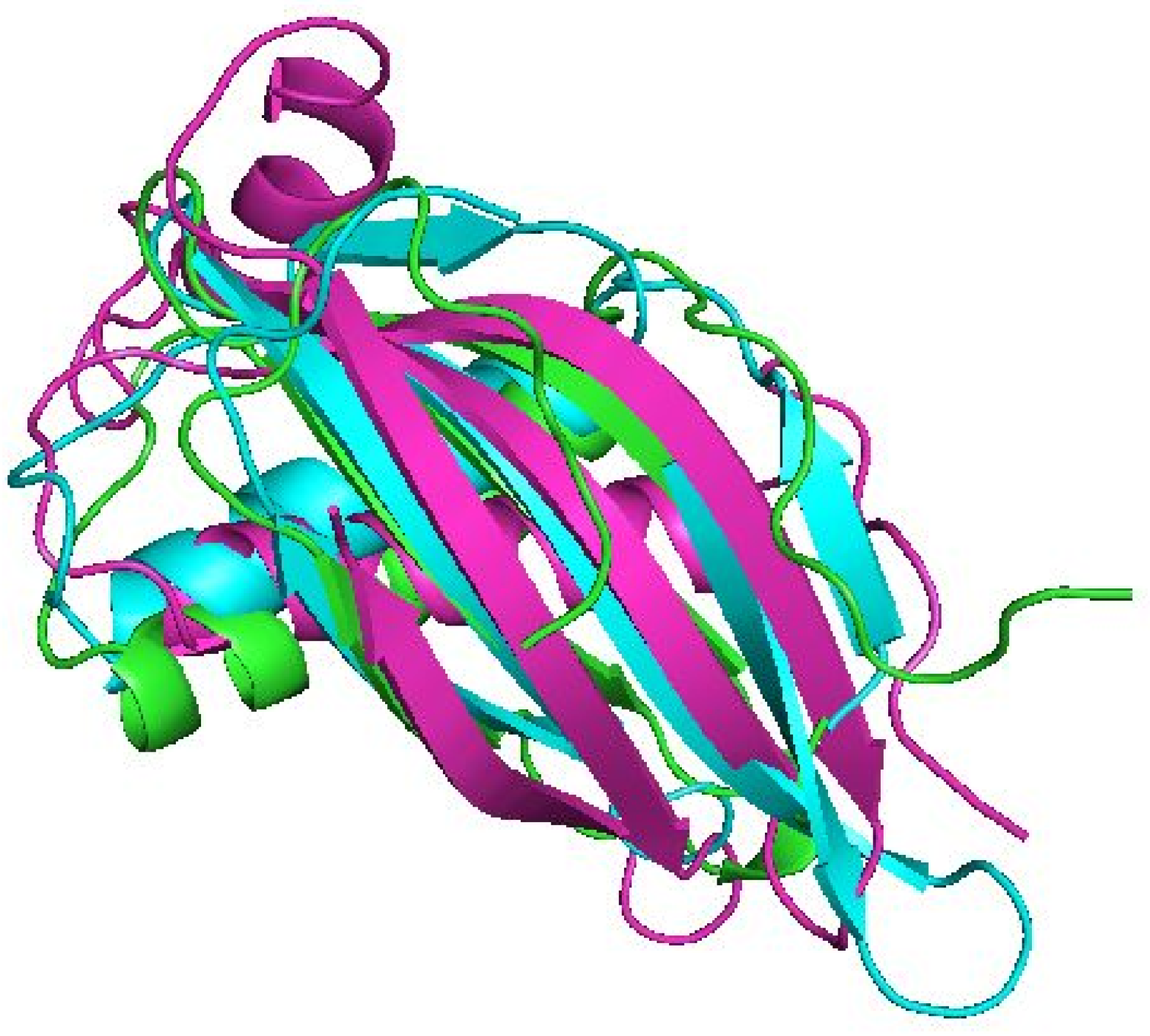}
  \caption{Set 8}\label{}
\endminipage\hfill
\minipage{0.32\textwidth}
	\includegraphics[width=\linewidth]{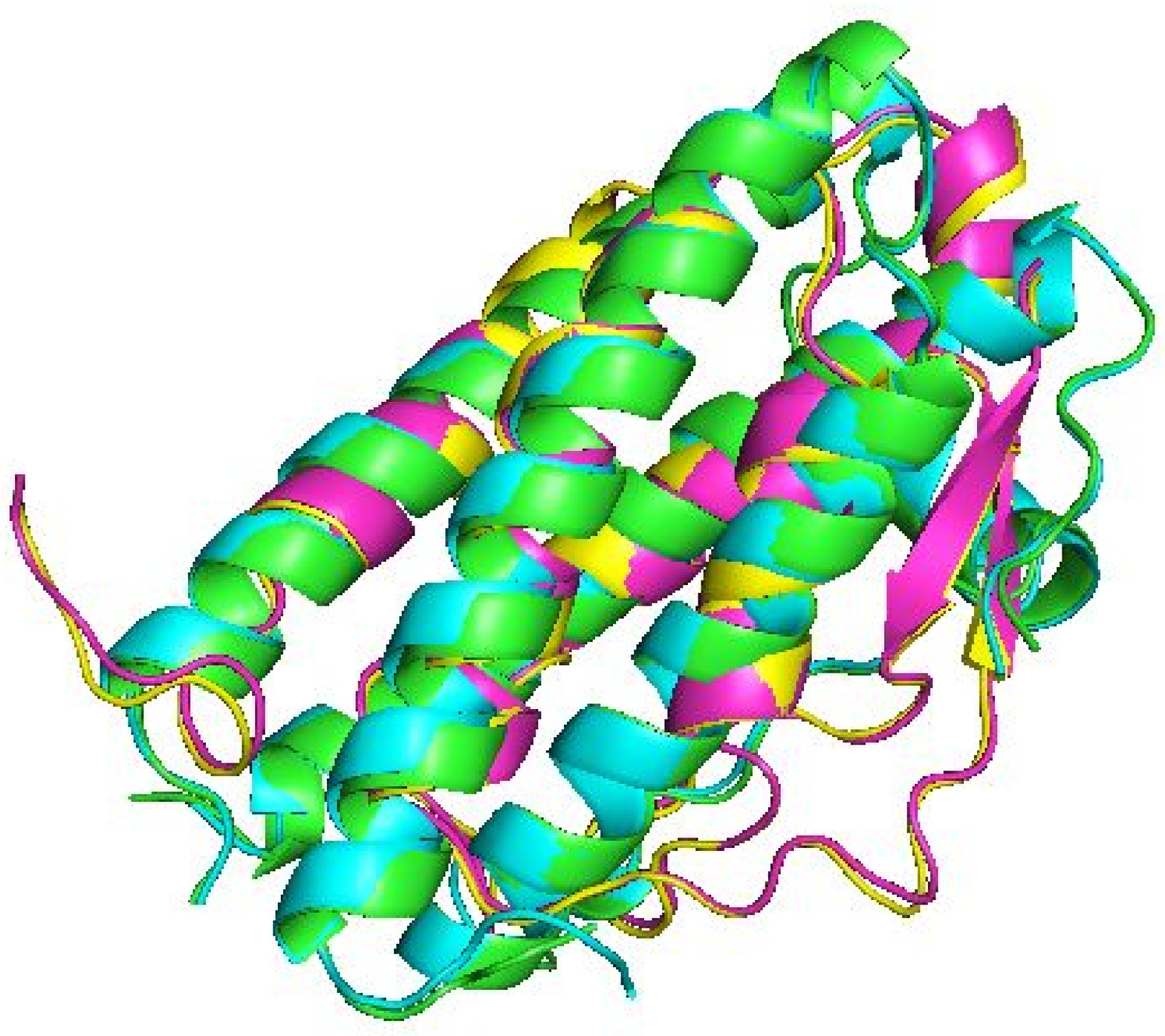}
  \caption{Set 8}\label{}
\endminipage\hfill
\minipage{0.32\textwidth}
	\includegraphics[width=\linewidth]{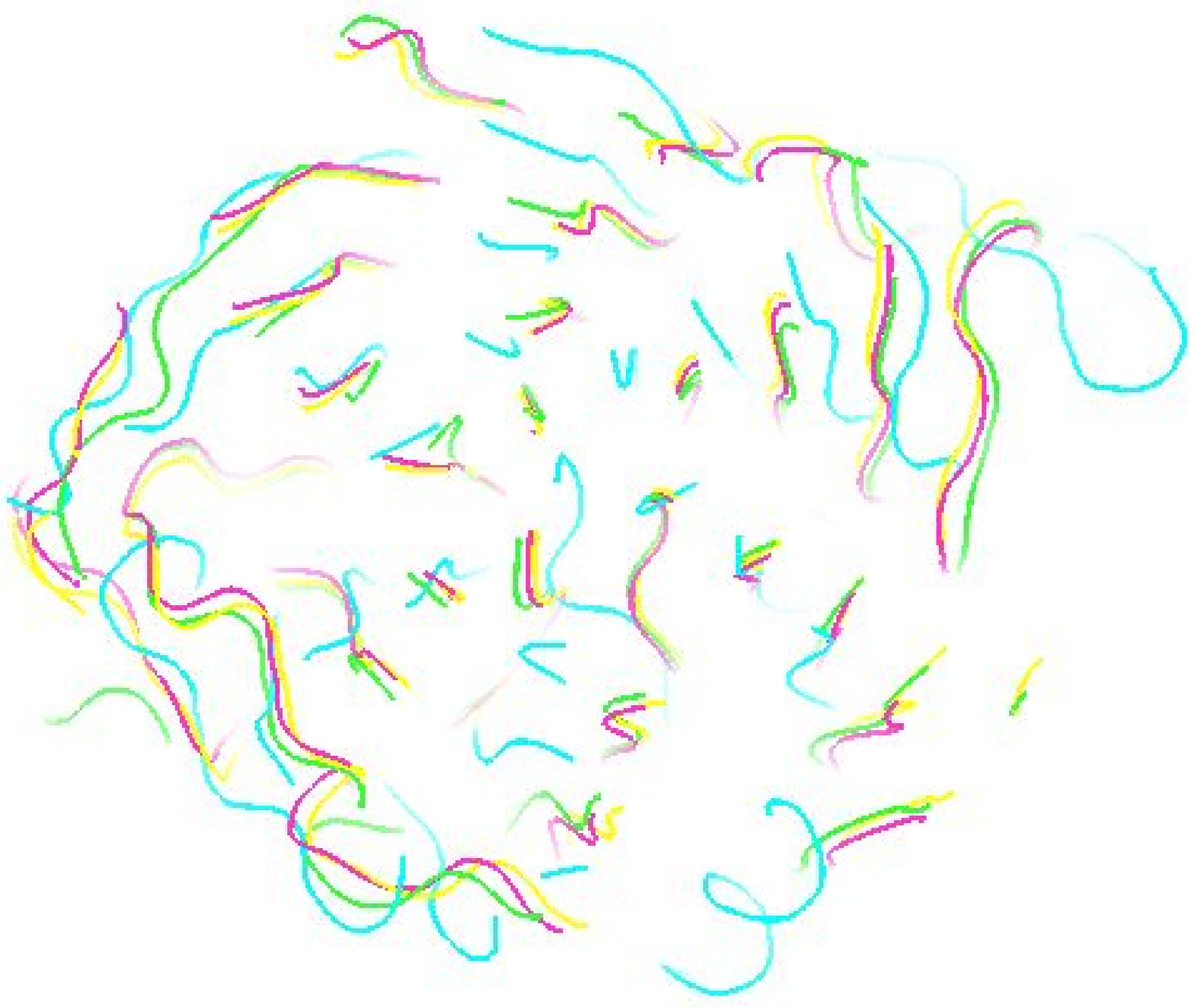}
  \caption{Set 10 LIPR}\label{}
\endminipage\hfill
\end{figure}\\

The above 3 sets have been chosen, after numerous trials, for their significantly low sequence similarity. The motive is to show that proteins that would never have been labeled as similar, even by the most powerful MSA techniques, can be aligned using MASCOT. This is possible because the sequential representation used here consists of SSE elements and not primary residues. Set 8, 9, and 10 represent three bands of sequence identity within the twilight zone. They have rmsd of \texttt{3.61}\AA{}, \texttt{0.1}\AA{}, and \texttt{3.15}\AA{} respectively.

\subsection{Pig, Malaria, Human, and Dogfish - connected?}
The 'Tree of life' has sprung many branches over millennia. Could the branches for pigs, malarial parasites, humans, and dogfish have had a common root at some point of time? The structures below have been taken from these species and an alignment is sought to gain more insight:
\begin{table}[h]
\caption{\label{}The table shows the sets used in this section:} 
\begin{center}
\begin{tabular}{|c|p{10cm}|c|c|}
\hline
   
 \hspace{1mm} Name \hspace{1mm} & \hspace{1mm} PDB ids \hspace{1mm} & \hspace{1mm} Count \hspace{1mm} & \hspace{1mm} T \hspace{1mm}\\ \hline
  
Set 11 & 1MLD:A 1MLD:B 1MLD:C 1MLD:D 1T2D:A 1I0Z:A 1I0Z:B 1LDM:A &  8 &  49s  \\ \hline
\end{tabular}
\end{center}
\end{table}
\begin{figure}[htb]
\minipage{0.5\textwidth}
	\includegraphics[width=\linewidth]{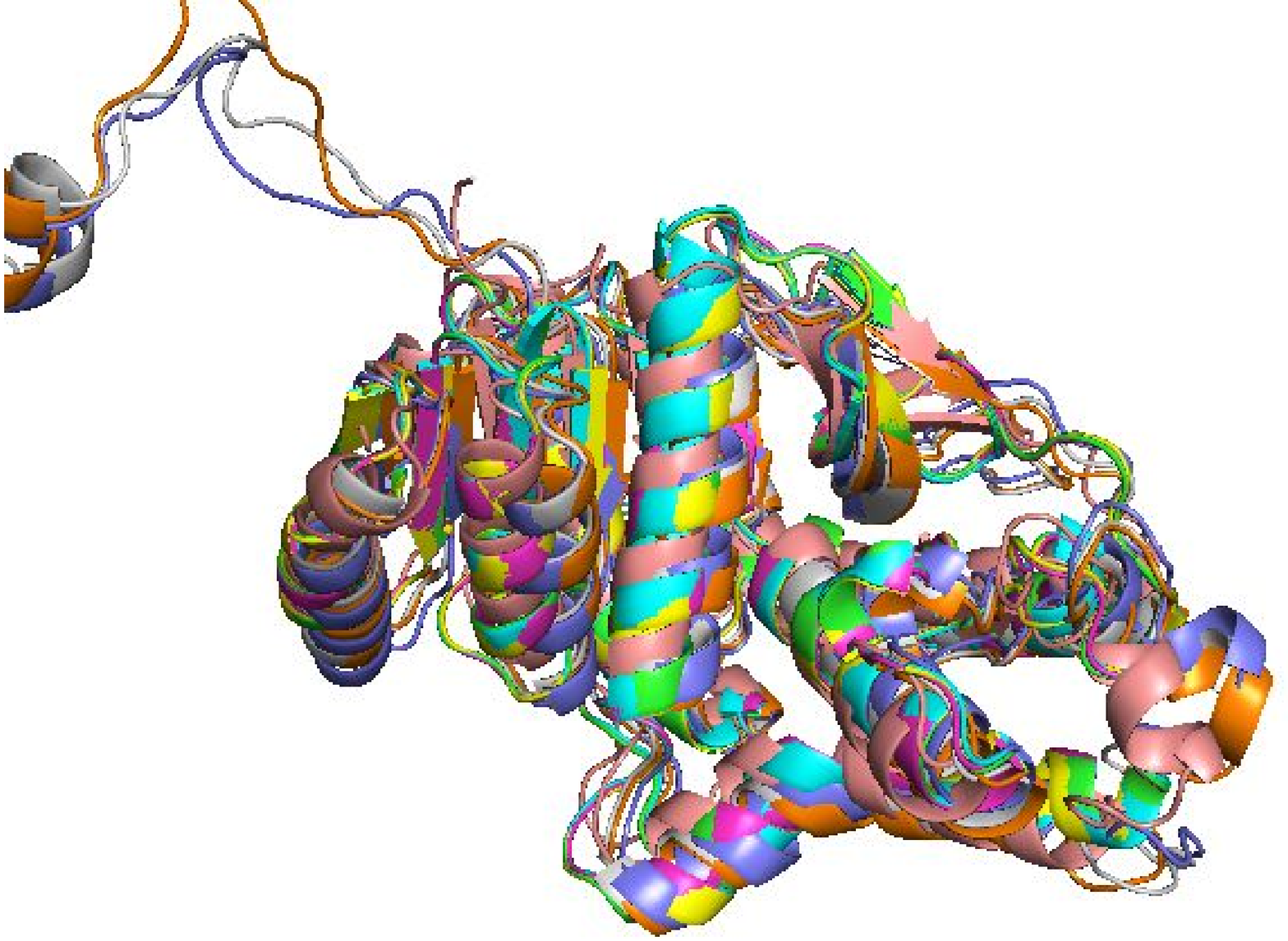}
  \caption{Set 11}\label{}
\endminipage\hfill
\minipage{0.5\textwidth}
	\includegraphics[width=\linewidth]{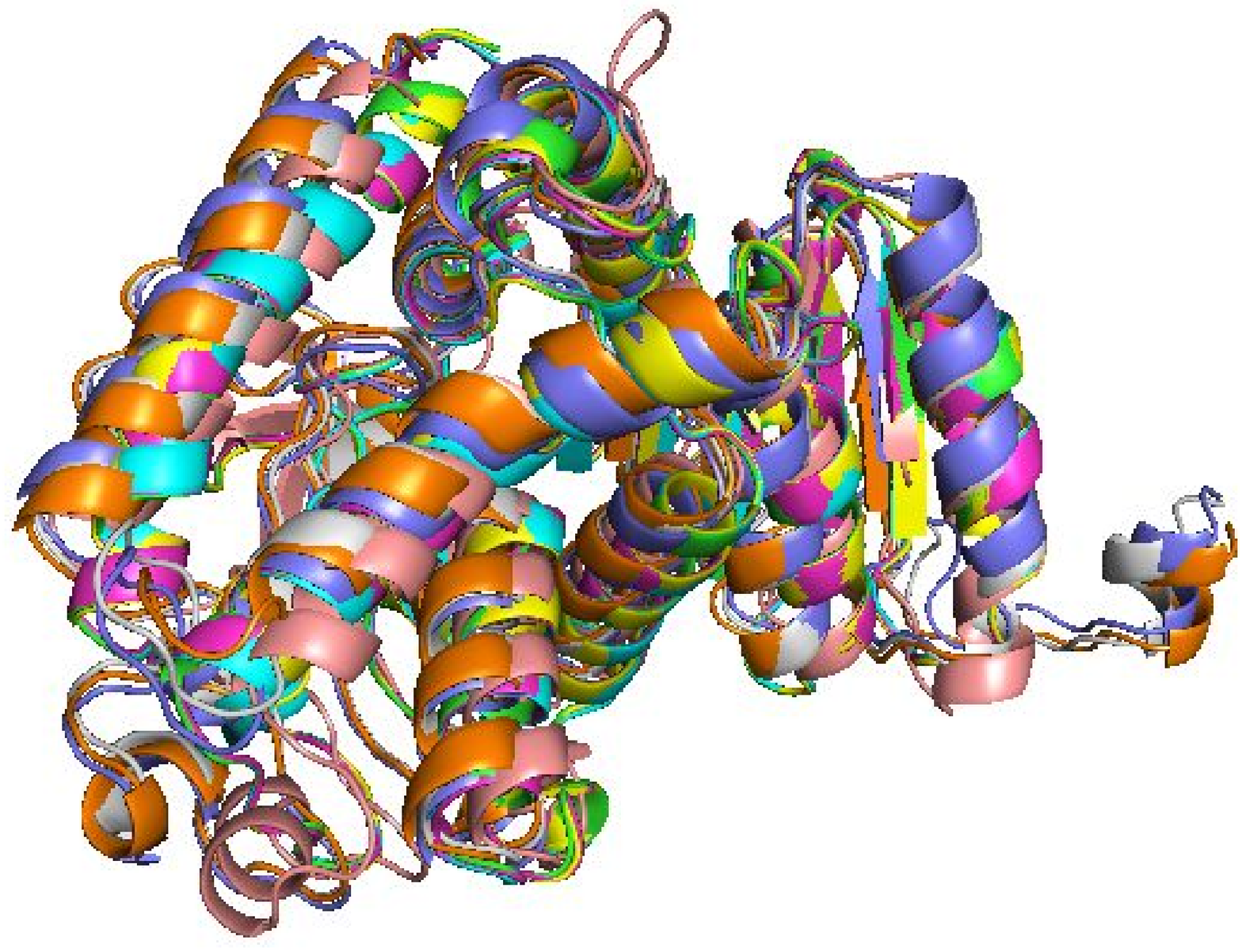}
  \caption{Set 11}\label{}
\endminipage\hfill
\end{figure}\\

The crystal structure of mitochondrial malate dehydrogenase from porcine heart (1MLD) contains four identical subunits\cite{Berman01012000}. Plasmodium falciparum, the causative agent of malaria, uses the protein 1T2D to enhance NAD+ regeneration. Incidentally this protein is being used for new anti-malarial drugs \cite{Berman01012000}. 1IOZ, a protein from Homo sapiens, is produced by the HRAS and HRAS1 genes \cite{Berman01012000}. 1LDM represents the crystal structure of M4 apo-lactate dehydrogenase from the spiny dogfish (Squalus acanthius)\cite{Berman01012000}. 

Figures 18 and 19 are the same alignment viewed from different angles. MASCOT finds striking similarities among these molecules with rmsd \texttt{2.885}\AA{}, indicating that at some point of time the branches for these species might indeed have had some common ancestor.

\subsection{Human, Chicken, Rabbit, Yeast, and Nematode}
An ensemble group of proteins have been taken from the species mentioned above. Could molecules taken from such diverse taxa be aligned to find structural similarity?
\begin{table}[h]
\caption{\label{}The table below shows the sets used in this section:} 
\begin{center}
\begin{tabular}{|c|p{10cm}|c|c|}
\hline
   
 \hspace{1mm} Name \hspace{1mm} & \hspace{1mm} PDB ids \hspace{1mm} & \hspace{1mm} Count \hspace{1mm} & \hspace{1mm} T \hspace{1mm}\\ \hline
  
Set 12 & 1SSG:A 1SSG:B 1HTI:A 1HTI:B 1R2S:A 1R2T:A 1MO0:A 1MO0:B 7TIM 3YPI &  10 &  47s  \\ \hline
\end{tabular}
\end{center}
\end{table}
\begin{figure}[H]
\makebox[\linewidth]{
\includegraphics[scale=0.4]{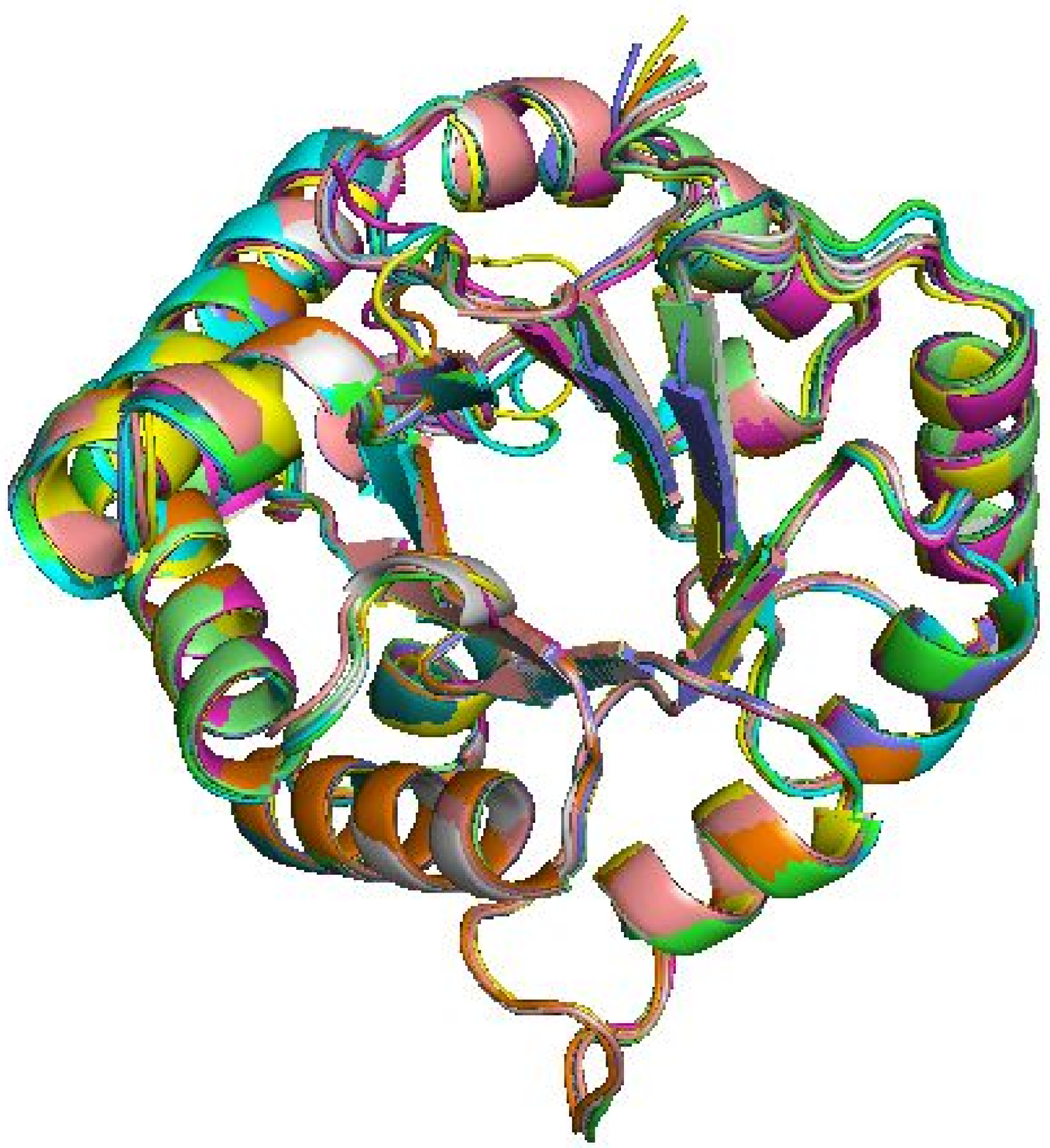}
}
\caption{Set 12}
\label{eqn}
\end{figure}

Different taxa perform the same function in their own way .For example, glycolysis is the 'metabolic pathway' \cite{romano1996evolution} using which glucose is broken down to form free energy. Chicken does this using the protein 1SSG \cite{Berman01012000}. Humans do the same thing using protein 1HTI\cite{Berman01012000} . We applied MASCOT to structures taken from rabbit muscle (1R2S, 1R2T), baker's yeast (7TIM, 3YPI), and nematode (1MO0), with an rmsd of \texttt{1.74}\AA{} to confirm that these proteins are used for the same purposes.

\subsection{Seafood allergy in Fish!}
Rats and humans are known to have allergy towards seafood. This is generally caused due to the presence of some proteins causing havoc in the immune system. Can such propensity be exhibited among fishes?
\begin{table}[h]
\caption{\label{}The table below shows the sets used in this section:} 
\begin{center}
\begin{tabular}{|c|p{10cm}|c|c|}
\hline
   
 \hspace{1mm} Name \hspace{1mm} & \hspace{1mm} PDB ids \hspace{1mm} & \hspace{1mm} Count \hspace{1mm} & \hspace{1mm} T \hspace{1mm}\\ \hline
  
Set 13 & 1RWY:A 1RJV:A 4CPV 3PAL 1BU3 5PAL &  6 &  5s  \\ \hline
\end{tabular}
\end{center}
\end{table}
\begin{figure}[H]
\makebox[\linewidth]{
\includegraphics[scale=0.4]{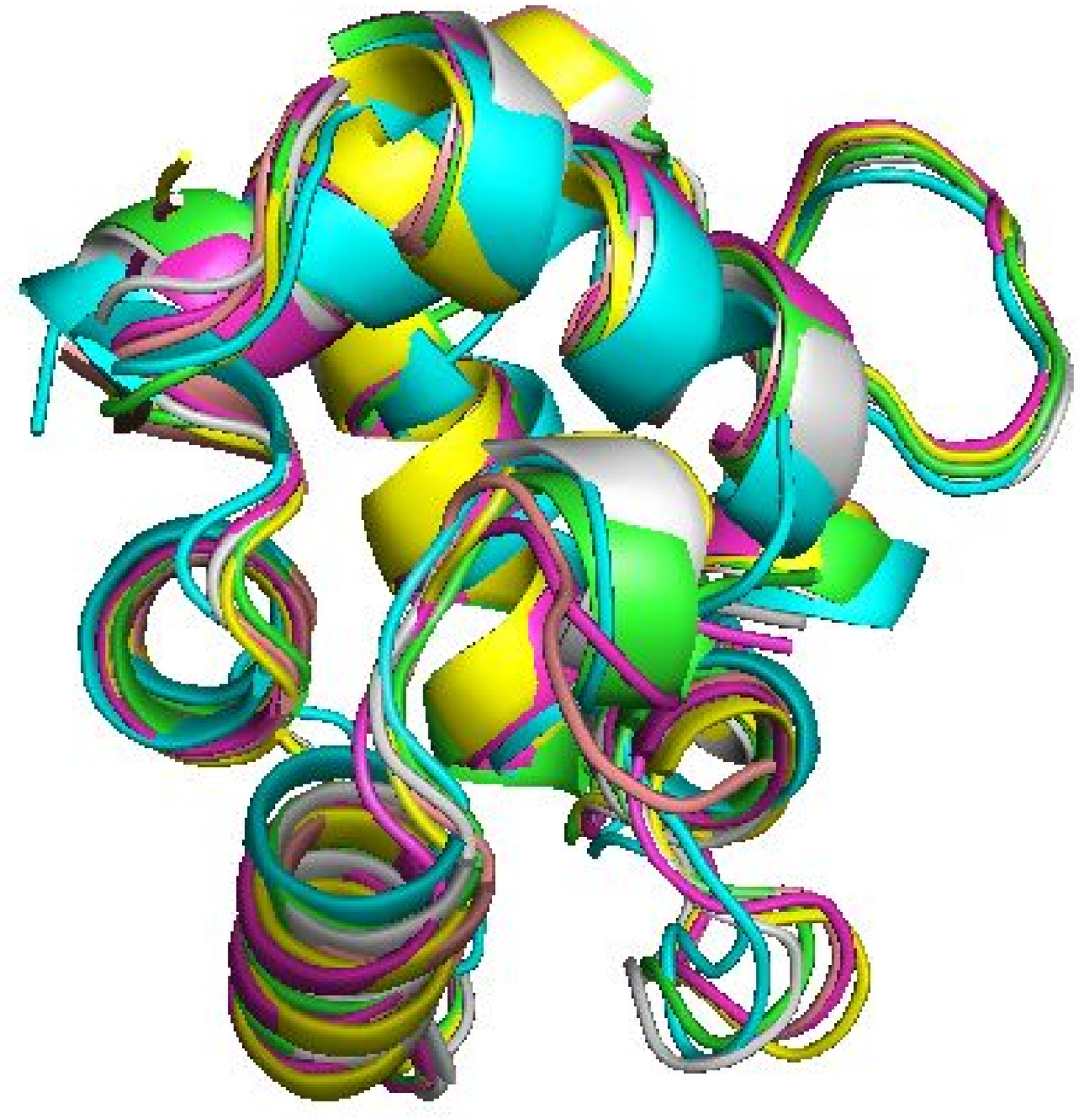}
}
\caption{Set 12}
\label{eqn}
\end{figure}
1RWY and 1RJV are known to cause seafood allergy in common brown rats and humans. After experimenting on a host of proteins we found out some fishes too have proteins with similar structure. For example, proteins 4CPV, 3PAL, 1BU3, and 5PAL subsequently taken from common carp, pike, silver hake, and leopard shark have highly similar tertiary structures. Could this be an indication that these proteins might cause seafood allergy in these fishes? It turns out that indeed they do. A recent study by Swoboda et al \cite{swoboda2002recombinant} suggests that parvalbumins, such as the ones taken above, are major cross-reactive fish allergen. Figure 21 shows how MASCOT correctly aligns the EF hand motifs in these proteins, albeit with an rmsd of \texttt{3.82}\AA{}. 

\subsection{Conclusions}
This paper contributes towards the goal of comparing more than two protein structures, and finding biologically relevant similarities within them. To this end we focused on using a new approach by reducing the complexity of the three dimensional structures into meaningful SSE elements, and adopting a center-star approach to arrive at equivalences.

We have introduced MASCOT, which has been designed to overcome the major hurdles of a multiple alignment by using a sum-of-pairs heuristic that associates all proteins with the one that is 'closest' to the others among the input set. 

The core of this work took the form of experiments. A representative set of results from these experiments have been presented in chapter 4. Sets 1 to 6 are standard data sets used by other published algorithms. MASCOT can efficiently align the proteins belonging to the globin, serpin, and tim barrel superfamilies. Set 7 represents data taken from a gold standard database (SCOP), which is a sort of litmus test for MStA methods. Sets 8, 9, and 10 show how MASCOT totally ignores the primary sequence and finds common motifs in spite of low sequence identity. Interesting observations have been noted through sets 11, 12, and 13. For example, set 11 shows that protein structures across these species have been conserved. So, during creation of phylogenetic trees based on structure, MASCOT can be used to process subsets of proteins as sub-problems, which later combine leading up to a tree. Set 12 and 13 are classic cases of structure-function association that suggest MASCOT results could be used to find yet-unknown biological similarities through computational methods.

\section{Future Work}
\label{sect:future}
MASCOT can be extended to include the following functionalities in future:
\begin{enumerate}
  \item Improve accuracy for aligning theoretical proteins.
  \item Take advantage of the flexible nature of proteins and account for them when considering similarities.
  \item Derive a core structure from the input proteins for use as template for protein threading.
\end{enumerate}

\bibliographystyle{plain}
\bibliography{references}
\end{document}